\documentclass[aps,prd,amsmath,amssymb,nofootinbib,longbibliography,noeprint,10pt,twocolumn,eqsecnum]{revtex4-2}
\usepackage{hyperref}
\usepackage{amsthm} 
\usepackage{dutchcal}
\usepackage{color}
\usepackage[dvipsnames]{xcolor}
\usepackage{mathtools}
\usepackage{mathrsfs}
\usepackage{bbm}
\usepackage{enumitem}
\usepackage{comment}
\usepackage[normalem]{ulem}
\usepackage{hyperref}
\usepackage{dcolumn}

\newcommand{\Tr}{\operatorname{Tr}}
\newcommand{\dd}{{\rm d}}

\newcommand{\eq}[1]{\eqref{#1}}
\newcommand{\BA}{\mathbf{A}}
\newcommand{\BB}{\mathbf{B}}
\newcommand{\BC}{\mathbf{C}}

\newcommand{\BS}{\mathbf{S}}
\newcommand{\BE}{\mathbf{E}}

\newcommand{\BG}{\mathbf{G}}
\newcommand{\BH}{\mathbf{H}}
\newcommand{\BI}{\mathbf{I}} 
\newcommand{\BK}{\mathbf{K}} 
\newcommand{\BL}{\mathbf{L}}
\newcommand{\BM}{\mathbf{M}}
\newcommand{\BN}{\mathbf{N}}
\newcommand{\BQ}{\mathbf{Q}}
\newcommand{\BU}{\mathbf{U}}
\newcommand{\BV}{\mathbf{V}}
\newcommand{\BX}{\mathbf{X}}
\newcommand{\BY}{\mathbf{Y}}
\newcommand{\BDelta}{\mathbf{\Delta}}

\newcommand{\NP}{\mathrm{NP}}
\newcommand{\NPMV}{\mathrm{NPMV}}
\newcommand{\NPCC}{\mathrm{NPCC}}
\newcommand{\DF}{\mathrm{DF}}
\newcommand{\DFCC}{\mathrm{DFCC}}
\newcommand{\CN}{\mathcal{CN}}
\newcommand{\HN}{H_N}
\newcommand{\HS}{H_S}
\newcommand{\PN}{P_N}
\newcommand{\PS}{P_S}
\newcommand{\ch}{\mathcal{h}}
\newcommand{\barF}{{\bar F}}

\newcommand{\bs}{\!\!\!\!\!} 
\newcommand{\lb}{\!\!\!}

\newcommand{\substantialchange}[1]{#1}

\newcommand{\diag}{{\rm diag}\,}
\newcommand{\Diag}{{\rm Diag}\,}

\def\be{\begin{equation}}
\def\ee{\end{equation}}
\def\ba{\begin{aligned}}
\def\ea{\end{aligned}}
\def\ben{\begin{enumerate}}
\def\een{\end{enumerate}}
\def\bi{\begin{itemize}}
\def\ei{\end{itemize}}

\newcommand{\MSE}{\mathrm{MSE}}
\newcommand{\ME}{\mathrm{ME}}
\newcommand{\trap}{{\rm trap}}

\begin{document}

\title{\bf Optimal robust detection statistics for pulsar timing arrays}
\date{today}

\author{Rutger van Haasteren}  
\email{rutger@vhaasteren.com}
\affiliation{Max-Planck-Institut f{\"u}r Gravitationsphysik (Albert-Einstein-Institut), Callinstra{\ss}e 38, D-30167 Hannover, Germany\\
Leibniz Universit{\"a}t Hannover, D-30167 Hannover, Germany}

\author{Bruce Allen} 
\email{bruce.allen@aei.mpg.de}
\affiliation{Max-Planck-Institut f{\"u}r Gravitationsphysik (Albert-Einstein-Institut), Callinstra{\ss}e 38, D-30167 Hannover, Germany\\
Leibniz Universit{\"a}t Hannover, D-30167 Hannover, Germany}

\author{Joseph D.\ Romano}
\email{joseph.romano@utrgv.edu}
\affiliation{Department of Physics and Astronomy, University of Texas Rio Grande Valley,\\
One West University Boulevard, Brownsville, TX 78520, USA
}

\begin{abstract}
  \noindent
  Pulsar timing arrays (PTAs) seek to detect a nano-Hz stochastic
  gravitational-wave background (GWB) by searching for the
  characteristic Hellings and Downs angular pattern of timing residual
  correlations. So far, the evidence remains below the conventional
  $5$-$\sigma$ threshold, as assessed using the literature-standard
  ``optimal cross-correlation detection statistic''.  While this quadratic
  combination of cross-correlated data maximizes the {\em deflection}
  (signal-to-noise ratio), it does not maximize the detection
  probability at fixed false-alarm probability (FAP), and therefore is
  not Neyman-Pearson (NP) optimal for the assumed noise and signal
  models.  The NP-optimal detection statistic is a different quadratic
  form, but is not used because it also incorporates autocorrelations,
  making it more susceptible to uncertainties in the modeling of
  pulsar timing noise.  Here, we derive the best compromise: a
  quadratic detection statistic which is as close as possible to the
  NP-optimal detection statistic (minimizing the variance of its
  difference with the NP statistic) subject to the constraint that it
  only uses cross-correlations, so that it is less affected by pulsar noise
  modeling errors. We study the performance of this new $\NPMV$ statistic
  for a simulated PTA whose noise and (putative) signal match those of
  the NANOGrav 15-year data release: GWB amplitude $A_{\rm
    gw}=2.1\times 10^{-15}$ and spectral index $\gamma=13/3$.
  Compared to the literature-standard ``optimal" cross-correlation 
  detection statistic,
  the $\NPMV$ statistic increases the detection probability by $47\%$
  when operating at a $5$-$\sigma$ FAP of $\alpha = 2.9 \times
  10^{-7}$.
\end{abstract}


\maketitle

\section{Introduction}
\label{s:intro}

Four pulsar timing array (PTA) collaborations have reported evidence for a nano-Hz gravitational-wave background
(GWB)~\citep{agazieNANOGrav15Yr2023b,reardonSearchIsotropicGravitationalwave2023,antoniadisSecondDataRelease2023e,xuSearchingNanoHertzStochastic2023}. The
time of arrival fluctuations from the pulsars that they monitor
display a characteristic pattern of angular correlations among
different pulsars, first predicted by Hellings \& Downs
(HD)~\citep{hellingsUpperLimitsIsotropic1983,allenVarianceHellingsDownsCorrelation2023a}.
One potential source for this GWB is a population of supermassive
black-hole binaries at the centers of galaxies
\citep{agazieNANOGrav15Yr2023d,antoniadisSecondDataRelease2024a,antoniadisSecondDataRelease2024}. Other,
more exotic sources have also been proposed and
studied~\citep{afzalNANOGrav15Yr2023}.

While they have exhibited several types of evidence in favor of the
GWB, none of the PTAs has made a detection claim.  To do so,
PTAs must demonstrate that their data are consistent with the HD
pattern with at least a $5$-$\sigma$-equivalent false-alarm probability
(FAP)~\citep{allenInternationalPulsarTiming2023}.  For this purpose,
the PTA community has adopted the so-called ``optimal cross-correlation
detection statistic''~\citep{anholmOptimalStrategiesGravitational2009,
  chamberlinTimedomainImplementationOptimal2015,vigelandNoisemarginalizedOptimalStatistic2018,
  allenHellingsDownsCorrelation2023a,allenOptimalReconstructionHellings2024}. Here, we call this the
\emph{cross-correlation deflection} statistic, indicated with the subscript ``$\DFCC$".

The cross-correlation deflection statistic is an example of a quadratic 
{\it detection statistic} (also known as a {\it test statistic})
defined by the quadratic form
\begin{equation}
  D(z\, |\, \BQ) \equiv z^\dagger \BQ\, z\,.
\label{e:quad_DS}
\end{equation}
Here, $z$ denotes the data (a complex column vector of dimension $n$,
with $\dagger$ denoting complex-conjugate transpose) and $\BQ$ denotes
the {\it filter} (or {\it kernel}) of the quadratic form, which is a
Hermitian matrix.  From here forward, we will call $D$ a ``statistic''
and $\BQ$ a ``filter''.  

In general, a detection statistic can be an {\it arbitrary} function
of the data, $D\equiv D(z)$: those defined by a quadratic form
\eqref{e:quad_DS} are thus a restricted class of detection statistics.
In what follows, we will omit the argument $\BQ$ from the statistic if
a statement or definition that we make applies to a general detection
statistic.
To emphasize that a particular formula or definition is specific to
quadratic detection statistics, then we will include the argument
$\BQ$ as in $D(z|\BQ)$.

In the simplest model, the dimension of $z$
is the number of pulsars, and each component of $z$ is the complex
Fourier amplitude of the timing residual for that pulsar at one
frequency. In more realistic models, $z$ has several complex Fourier
amplitudes per pulsar, typically grouped together by pulsar.

A detection statistic $D(z)$ together with a real {\it threshold}
$\tau$ define a {\it decision rule} for either accepting or 
rejecting the ``null'' hypothesis $\HN$.  
If $D(z)<\tau$ then $\HN$ is accepted: no GWB was found.  
If $D(z)>\tau$ then $\HN$ is rejected: a GWB was detected.  
The threshold $\tau$ is selected so that the probability 
$\alpha$ of claiming detection of a GWB when no GWB was present
has a small value, e.g., $\alpha= 2.9\times 10^{-7}$ for 5-$\sigma$
(Gaussian-equivalent) significance.  

The null hypothesis $\HN$ is equivalent to providing a probability
distribution $\PN(z)\equiv P(z|\HN)$ on the data $z$.  That allows us
to define ensemble averages (or expectation values) of functions 
$f(z)$:
\be
\langle f(z)\rangle_{\HN}\equiv \int \dd^n z\> f(z) \PN(z)\,.
\label{e:<>}
\ee
We will also make use
of a \emph{signal hypothesis} $\HS$ with its corresponding probability
distribution $\PS(z)\equiv P(z|\HS)$.  Expected values with respect to 
that probability distribution are given by~\eqref{e:<>} with $\HN$ replaced
by $\HS$ and $\PN$ replaced by $\PS$.

The {\it false alarm probability} (FAP) $\alpha$ is the probability of
the detection statistic exceeding the detection threshold $\tau$ in
the \emph{absence} of a GWB signal, i.e., under the null
hypothesis. It is
\be
\alpha \equiv {\rm Prob}(D>\tau|\HN)
\equiv \bigl\langle \theta(D(z) - \tau) \bigr\rangle_{\HN} \, ,
\label{e:FAP}
\ee
where $\theta(x)$ is the Heaviside step function
\substantialchange{
\begin{equation}
  \label{e:stepfn}
  \theta(x) =
  \begin{cases}
    0 & \text{if } x < 0 \, , \\
    \frac{1}{2} & \text{if } x = 0 \, , \\
    1 & \text{if } x > 0 \, .
  \end{cases}
\end{equation}
From $\theta(x)+\theta(-x) = 1$ and $\bigl\langle 1 \bigr\rangle = 1$
it follows that $1-\alpha = \bigl\langle \theta(\tau - D(z))
\bigr\rangle_{\HN}$.
}

\substantialchange{
The {\em detection probability} $1-\beta$ is the
probability of the detection statistic exceeding the threshold $\tau$
in the {\it presence} of a GWB signal. It is
\be
1-\beta \equiv {\rm Prob}(D>\tau|\HS) \equiv
\bigl\langle \theta(D(z) - \tau) \bigr\rangle_{\HS} \,,
\label{e:DP}
\ee
where the expected value uses the signal hypothesis $\HS$.  The
\emph{false dismissal probability} $\beta = \bigl\langle \theta(\tau -
D(z)) \bigr\rangle_{\HS}$ is the probability that a GWB signal
produces a statistic value which is too small to exceed the detection
threshold $\tau$.
}

The filter $\BQ$ for the general deflection statistic is chosen to
maximize the ratio of the square of the difference of the expected
values of $D(z|\BQ)$ under the signal and noise hypotheses $\HS$ and
$\HN$, respectively, to its variance under $\HN$: this is
sense in which it is ``optimal''.
Surprisingly, even for the
case of two simple hypotheses (models with no free
parameters\footnote{In practice, for composite hypotheses (models
with free parameters), these statistics are still 
used~\citep{agazieNANOGrav15Yr2024,vallisneriPosteriorPredictiveChecking2023,vanhaasterenCalculationPvaluesQuadratic2025}.}), the deflection-optimal statistic is
\emph{not} the Uniformly Most Powerful (UMP) test.

Among all possible statistics [arbitrary functions $D(z)$] the
UMP test is guaranteed to maximize the detection probability at fixed
FAP. It may be obtained via the Neyman-Pearson (NP) Lemma, 
and is any monotonic function of the ratio of the probability
distributions defining $\HS$ and $\HN$, see~\eqref{e:QNP_def}.  If, as
in this paper, $\HS$ and $\HN$ are multivariate zero-mean Gaussian
distributions, then the statistic is a quadratic form as in
\eqref{e:quad_DS}, see \eqref{e:ump_filter}.  In this paper, the
UMP statistic and corresponding filter are indicated with the
subscript ``NP''.  They are not the same as (or proportional to, or
monotonic functions of) the $\DFCC$ statistic and corresponding filter.

Why has the PTA community chosen to adopt a statistic ($\DFCC$) which
is not the most powerful one?  The reason is that the $\DFCC$ statistic
depends only upon pulsar timing residual cross-correlations, and not
upon pulsar timing residual autocorrelations, whereas the $\NP$
statistic depends upon \emph{both}. That is undesirable, because it
makes the $\NP$ statistic more vulnerable to errors arising from modeling
of the (non-GWB) pulsar noise. To indicate this weakness, we say that
the $\NP$ statistic is \emph{not robust}.

In this paper, we exploit the gap between the $\DFCC$ and $\NP$
statistics.  We construct a quadratic statistic, whose performance
lies in between $\NP$ and $\DFCC$, but which is robust.  More
precisely, we find the filter $\BQ$ whose quadratic statistic
\eqref{e:quad_DS} is as close as possible (in the minimum-variance
sense) to the $\NP$ statistic under the null hypothesis, but which is
robust because it does not use auto-correlations. This corresponds to
filters $\BQ$ which vanish on the diagonal.  At a given FAP, we will
see that the detection probability for this
``Neyman-Pearson-Minimum-Variance'' ($\NPMV$) statistic is higher than
that of the conventional $\DFCC$ statistic.

We employ a standard technique to characterize and compare different
statistics: via their receiver operating characteristic (ROC) curves
(see Secs.~\ref{ss:ROC_toy} and~\ref{ss:ROCcurves}).  These plot the
detection probability as a function of FAP.  We use these to show that
in realistic cases the $\NPMV$ statistic has better performance
(larger area under the ROC curve) than the standard $\DFCC$ statistic.
For a PTA similar to the published NANOGrav 15-year dataset, the
$\NPMV$ statistic increases the detection probability by $\sim\!47\%$
when operating at the $5$-$\sigma$ detection threshold FAP of $2.9
\times 10^{-7}$ (see Fig.~\ref{fig:ROC_NG15}).

Given its improved performance, we believe that the $\NPMV$ detection
statistic should replace the (currently standard) ``optimal" 
cross-correlation statistic for evaluating detection claims and computing
$p$-values in PTA data analysis.%
\footnote{\label{fn:DefinePvalue}The $p$-value for a detection
statistic is defined as $p\equiv {\rm Prob}(D>D_{\rm obs}|\HN)$, where
$D_{\rm obs}\equiv D(z_{\rm obs})$ for the observed data $z_{\rm
  obs}$.}  \substantialchange{At a given FAP, the $\NPMV$ statistic achieves higher
detection probability while remaining robust to pulsar noise modeling
errors, the latter demonstrated in Sec.~\ref{s:mismodel} for a simple
toy model.  We also include an Appendix, where we derive analytic
expressions for the mean (noise-mismodeling) error and mean-squared
error in the $p$-value.  These are valid when the pulsar-pulsar
correlations are well-approximated by a multivariate Gaussian.}

To encourage its adoption, we provide a Python implementation of the
$\NPMV$ filter and statistic in the \texttt{enterprise\_extensions}
package~\citep{ellisENTERPRISEEnhancedNumerical2020,taylorEnterprise_extensions2021}.

\section{Autocorrelations and cross-correlations}
\label{s:blocks}

To facilitate discussion of auto- and cross-correlations, assume
(without loss of generality) that the vector $z$ of data is
\emph{grouped by pulsar}.  This means that the first $m_1$ components
are data from pulsar $1$, the next $m_2$ components are data from
pulsar $2$, and so on.  For each given pulsar, these data might be
Fourier amplitudes at different frequencies, or different amplitudes
in a Legendre polynomial decomposition, or timing residuals measured
at different epochs.

Matrices such as the covariance $\Sigma$, whose rows and
columns carry the same indices as $z$, may then be described as being
composed of \emph{blocks}.  These blocks are delineated by drawing
vertical lines between any neighboring columns associated with
different pulsars, and by drawing horizontal lines between any
neighboring rows associated with different pulsars.

The \emph{block diagonal} of such a matrix consists of those
\emph{blocks} containing at least one diagonal element of the matrix.

For such a matrix $\BM$, we define a linear operator $\Diag\!$ as
follows: $\Diag \BM$ is the matrix obtained from $\BM$ by zeroing all
entries that lie \emph{off} the block diagonal.  This differs from
$\diag \BM$, which is obtained from $\BM$ by setting all non-diagonal
entries to zero.  (If there is only one data sample per pulsar, then
$\Diag \BM = \diag \BM$.)  A matrix $\BM$ is said to be {\it
  block-diagonal} iff $\Diag \BM = \BM$.

From here forward, we will assume that the data vector $z$ is grouped
by pulsar. This means that the \emph{diagonal blocks} of covariance
matrices and quadratic filters contain autocorrelation terms.  The
remaining blocks, which contain cross-correlation terms, are called
the \emph{off-diagonal blocks}.

\section{Definitions of the detection statistics}
\label{s:definitions}

Here, we define the five different quadratic detection statistics
($\DF$, $\DFCC$, $\NP$, $\NPCC$, and $\NPMV$) considered in this
paper, and give formulas for four of the corresponding filters.  The
$\NPMV$ filter is derived in Sec.~\ref{s:NPW-derivation} and
derivations of the $\NP$ and $\DF$ filters are given
in~\cite{vanhaasterenCalculationPvaluesQuadratic2025}.

In what follows, the noise and signal hypotheses $\HN$ and $\HS$ are
assumed to be zero-mean multivariate Gaussian distributions for the
complex-valued data $z$.  These distributions are defined by the
covariance matrices $\BN$ and $\BS$, respectively, which are Hermitian
and positive-definite.  Later, starting in Sec.~\ref{s:CURN}, we will
restrict $\BN$ to the form most often used in PTA analyses.  But here,
$\BN$ and $\BS$ are \emph{arbitrary} Hermitian, positive-definite
matrices.

\subsection{Quadratic filter definitions}
\label{s:Q_defs}

As described in Sec.~\ref{s:intro}, the filter $\BQ$ for the general
deflection-optimal statistic is defined by the requirement that it
maximize the ratio of the square of the difference of the expected
values of $D(z|\BQ)$ under $\HS$ and $\HN$, respectively, to its
variance under $\HN$. Thus,
\begin{equation}
  \label{e:QDF_def}
  \BQ_\DF \equiv
  \mathop{\arg\max}_{\BQ} 
  \frac{\left(\langle D(z\,|\,\BQ)\rangle_{\HS} - \langle D(z\,|\,\BQ)\rangle_{\HN}\right)^2}
  {{\rm Var}[D(z\,|\,\BQ)]_{\HN}}\,,
\end{equation}
where the variance of (complex) $w$ is
\begin{equation}
{\rm Var}[w] \equiv \left\langle|w-\langle w\rangle|^2\right\rangle
= \left\langle |w|^2 \right \rangle - \left|\left\langle w \right\rangle\right|^2 \, .
\end{equation}
As shown by~\cite{vanhaasterenCalculationPvaluesQuadratic2025},
requirement~\eqref{e:QDF_def} leads to:
\begin{equation}
\label{e:def_filter}
\BQ_\DF =\BN^{-1}\left(\BS - \BN\right)\BN^{-1}\, .
\end{equation}
For arbitrary covariance matrices $\BN$ and $\BS$,
$\BQ_\DF$ does not vanish on the block diagonal, so in general it
uses both auto- and cross-correlations. [For the PTA CURN null
hypothesis, it only uses cross-correlations, see \eqref{e:CURN}
and Sec.~\ref{s:CURN}.]

A {\it robust} version of the deflection-optimal statistic, which uses
only cross-correlations, can be derived in the same fashion as the
general $\DF$ statistic.  One first restricts the filter to use only
cross-correlations, then maximizes, obtaining
\be
\BQ_\DFCC = \BQ_\DF - \Diag \BQ_\DF \,.
\label{e:defcc_filter}
\ee
$\BQ_\DFCC$ is the literature-standard ``optimal cross-correlation detection statistic'',
obtained from \eqref{e:def_filter} by removing the autocorrelations
from the general $\DF$ filter.  It is used to search for GWBs with
terrestrial GW detectors like LIGO, and for cross-correlation PTA
analyses with non-standard null hypotheses, as discussed in
Sec.~\ref{s:hypotheses}.

Equation \eqref{e:def_filter} does not include the usual
data-independent normalization factor
$(\Tr[(\BN^{-1}(\BS-\BN))^2])^{1/2}$ in the denominator of
\eqref{e:def_filter}. This factor is usually inserted to ensure that
the variance of $\BQ_\DF$ is unity under $\HN$. Here, however, we
compare the performance of different detection statistics.  This
performance is encoded in the ROC curves, which give the detection
probability as a function of the FAP, and are independent
of the filter normalization.

As discussed in Sec.~\ref{s:intro}, the $\NP$-optimal statistic
maximizes the detection probability for a given FAP,
\begin{equation}
  \label{e:QNP_def}
  D_\NP \equiv
  \mathop{\arg\max}_{D\,|\,{\rm Prob}(D>\tau|\HN)=\alpha}
  {\rm Prob}(D>\tau|\HS) \,.
\end{equation}
It is
proportional to the log-likelihood ratio $\log P(z|\HS) - 
\log P(z|\HN) = \log(\PS(z)/\PN(z))$. 
In general, this quantity is not a quadratic function
of $z$. However, for our specific choices of multivariate 
zero-mean Gaussian distributions for $\HN$ and $\HS$, 
it {\it is} quadratic, with filter
\begin{equation}
\label{e:ump_filter}
\BQ_{\rm NP} \equiv \BN^{-1} - \BS^{-1}\,.
\end{equation}
As before, this could be normalized by dividing by a factor of $(\Tr
[(\BS^{-1}(\BS-\BN))^2])^{1/2}$, but this serves no purpose here.
Interestingly, in the limit of weak GWB signals (relative to the
noise, meaning $\BS \approx \BN$), $\BQ_\NP$ reduces to $\BQ_\DF$.

While the $\NP$ filter in \eqref{e:ump_filter} is the best possible
detection statistic (in the Neyman-Pearson UMP sense), it is not used in practice in
the PTA community because $\Diag \BQ_\NP \ne 0$.  This means that the
detection statistic incorporates autocorrelations, which in turn are
subject to pulsar noise mismodeling uncertainties.  To obtain a robust
version of the NP-optimal test, we first sought a statistic that
maximizes the detection probability at fixed FAP, subject to the
constraint that it does not use autocorrelation information.  However,
if the number of pulsars is $\ge 3$, then no such statistic exists,
because the autocorrelations can be reconstructed from the
cross-correlations.

We then sought a statistic that was quadratic in the data, of the form
$D(z\,|\,\BQ)$ given in \eqref{e:quad_DS}, with a filter $\BQ$
selected to maximize the detection probability at fixed FAP, subject
to the constraint that $\Diag \BQ = 0$, so that the statistic uses
only cross-correlations.  We denote this filter by $\BQ_\NPCC$, but we
were not able to obtain an analytic closed form for it.  It can be
found numerically (see Sec.~\ref{ss:ROC_toy}) but the form of the
solution (meaning its matrix ``shape'', not just the overall
normalization) depends upon the FAP, which makes it cumbersome to
characterize and employ.

The $\NPMV$ statistic is a compromise between the
cross-correlation-only deflection-optimal statistic $\DFCC$ and the
auto+cross-correlation Neyman-Pearson-optimal statistic $\NP$. By
definition, it is ``as close as possible" to the $\NP$ statistic,
while using only cross-correlated data.  More precisely, it minimizes
the variance (MV) under $\HN$ of its difference with the $\NP$
statistic, subject to the constraint that the filter $\BQ$ vanish on
the block diagonal:
\begin{equation}
  \label{e:argmin}
  \BQ_\NPMV \equiv
  \mathop{\arg\min}_{\BQ \,|\, \Diag \BQ = 0} 
  {\rm Var}[D(z\,|\,\BQ) - D(z\,|\,\BQ_\NP)]_{\HN} \,.
\end{equation}
For arbitrary $\BN$, $\BQ_\NPMV$ has a unique solution 
determined by~\eqref{e:argmin} (see Sec.~\ref{s:NPW-derivation}),
but it does not have a very simple form.
However, for block-diagonal $\BN$,
the unique solution is
\be
\BQ_\NPMV = \BQ_\NP - \Diag \BQ_{\NP}
\quad (\text{assuming } \BN = \Diag \BN)\,.
\label{e:npw_filter}
\ee 
This corresponds to simply removing the autocorrelation terms 
from the full $\NP$-optimal filter.
See Sec.~\ref{s:NPW-derivation} for a derivation of this
relation.

With this in mind, the $\NPMV$ statistic can be thought of as the
\emph{bronze medalist}: it ranks third overall, after $\NP$ and
$\NPCC$.  However, as we show in Sec.~\ref{s:comparison}, 
the $\NPMV$ statistic significantly outperforms
the $\DFCC$ statistic.  A set of ROC curves,
comparing $\NP$, $\NPCC$, $\NPMV$, and $\DFCC$ for a 
simple model%
\footnote{For this model, $\DF$ and $\DFCC$ are identical.}
is presented in Sec.~\ref{ss:ROC_toy}, Fig.~\ref{fig:ROC_TOY}.

\section{Null (noise) and signal hypotheses}
\label{s:hypotheses}

In the previous section, the noise and signal hypotheses $\HN$ and
$\HS$ were defined by arbitrary Hermitian, positive-definite
matrices $\BN$ and $\BS$.  However, conventional PTA analyses use
hypotheses $\HN$ and $\HS$ that make certain assumptions regarding
the properties of the pulsars and the GW signals.  These restrict the
form of $\BN$ and $\BS$.  In this section, we describe some of these
restricted forms of $\BN$ and $\BS$, as we need this to understand
how to apply our proposed methods to realistic PTA data.

\subsection{Block-diagonal and CURN null hypotheses}
\label{s:CURN}

For most stochastic background analyses, including GWB searches using
ground-, space-, and galactic-based detectors like LIGO, LISA, and
PTAs, the null hypothesis $\HN$ is defined by a {\it block-diagonal}
covariance matrix $\BN$:
\be
\BN = \Diag\BN\,.
\label{e:blockdiag}
\ee
This means that ``detector noise'' is uncorrelated between different
detectors, or in the case of PTAs, between different pulsars.

It is also typical in PTA analyses to set the null hypothesis
covariance $\BN$ to agree with the signal covariance matrix $\BS$ on
the block diagonal:
\begin{align}
&\BN = \Diag(\BS)\,.
\label{e:CURN}
\end{align}
This $\HN$ is called the CURN hypothesis, meaning ``common
uncorrelated red noise"; 
see Sec.~4.2 of 
\cite{ellisEfficientApproximationLikelihood2013} where 
this choice of null hypothesis was first introduced.
With this choice, the signal hypothesis
$\HS$ differs from the null hypothesis $\HN$ only via the effects of the
off-diagonal blocks of $\BS$.  These off-diagonal blocks contain
information about the Hellings and Downs (HD) cross-correlations
induced by the GWB.

CURN is a conservative choice of $\HN$.  Rather than saying ``there is
no GWB'', it says ``there is a background, but not of the type
described by the general theory of relativity (GR)''.  The CURN
background contains common-spectrum autocorrelations in pulsar timing
fluctuations, but without the HD cross-correlations of GR.  If the
null hypothesis were ``no GWB'', then the diagonal blocks of $\BS$
would contain \emph{additional} contributions from the GWB that are
not present in $\BN$.  Choosing CURN as $\HN$ makes the analysis less
sensitive to pulsar noise modeling erorrs but also
makes it less sensitive to the GWB. Hence, it is more conservative
than the alternative of a ``no GWB'' $\HN$.

One consequence of taking $\HN$ to be CURN is that the DF-optimal
filter $\BQ_\DF$ given in \eqref{e:def_filter} involves only
cross-correlations, thus implying $\BQ_\DF = \BQ_\DFCC$ for CURN.

\subsection{Relating DF and NP filters for realistic PTA data}
\label{s:relatingDFNP}

Realistic PTA data consists of multiple samples per pulsar.
These data may be modeled using Gaussian
processes~\citep{rasmussenGaussianProcessesMachine2006,vanhaasterenNewAdvancesGaussianprocess2014}.
For computational efficiency, realistic analyses use reduced-rank
representations for the model components, obtained via a chain of
basis transformations that whiten (see below) and reduce 
the dimension of the
original data~\citep{vanhaasterenCalculationPvaluesQuadratic2025}.
Current analysis codes already compute the standard $\DF$ and
$\DFCC$ filters in that reduced-dimension whitened basis.  
Here, we show how the suitably whitened $\NPMV$ filter may 
be directly obtained from the whitened $\DF$ filter for a 
block-diagonal noise hypothesis $\HN$.

The whitened data are given by
\be
\label{e:white1}
\tilde{z} \equiv \BN^{-1/2}z\,.
\ee
These are said to be ``whitened'' because the covariance matrix of
$\tilde{z}$ is the identity matrix $\BI$.  The whitening filter
$\BN^{-1/2}$ is well defined, because (as described above, and also
true in the reduced basis) $\BN$ is block diagonal, with each block
being Hermitian and positive-definite.

Under this whitening change of basis, detection statistics are
invariant. For the quadratic detection statistics considered 
in this paper, this means that
\be
D(z\,|\,\BQ) = z^\dagger \BQ z = D(\tilde z\,|\,\tilde\BQ) = \tilde z^\dagger \tilde\BQ\tilde z \,.
\label{e:whiten2}
\ee Substituting \eqref{e:white1} into \eqref{e:whiten2}, we see that
the whitened version of the filter $\BQ$ is given by
\be \quad \tilde \BQ\equiv\BN^{1/2} \BQ
\BN^{1/2}\, .
\label{e:whiten3}
\ee
In this way, we use a \textasciitilde\  to denote both whitened
data and whitened filters.

From \eqref{e:whiten3}, the whitened version of the $\DF$
filter~\eqref{e:def_filter} is
\be
\label{e:white4}
\tilde\BQ_\DF \equiv \BN^{1/2}\BQ_\DF \, \BN^{1/2} =
\BA -\BI\,,
\ee
where $\BI$ is the identity matrix, and
\be
\BA\equiv \BN^{-1/2} \, \BS \, \BN^{-1/2}
\label{e:BA}
\ee
is the inverse of the whitened inverse-signal covariance 
$\BN^{1/2}\BS^{-1}\BN^{1/2}$.
From \eqref{e:whiten3}, the whitened version of the $\NP$ filter~\eqref{e:ump_filter} is
\be
\begin{aligned}
\tilde\BQ_\NP 
&\equiv
\BN^{1/2} \BQ_\NP \, \BN^{1/2} = \BI-\BA^{-1}\\
&=(\BA-\BI)\BA^{-1} = \tilde\BQ_\DF(\BI+\tilde\BQ_\DF)^{-1}\,,
\label{e:ump_filter_whitened} 
\end{aligned}
\ee
where the final equality follows from~\eqref{e:white4}.  This means
that the whitened $\NP$ filter can be easily constructed from the
whitened $\DF$ filter.

From the whitened $\NP$ filter, it is easy to construct the whitened
$\NPMV$ filter by removing the diagonal blocks:
\be
\tilde\BQ_\NPMV =  \tilde\BQ_\NP 
- \Diag \tilde\BQ_\NP\,.
\label{e:ump_filter_fromdef} 
\ee
Since $\BN^{1/2}$ is block-diagonal, this again corresponds to 
setting the diagonal blocks of the whitened NP-optimal filter to zero.

In a more realistic setting, this still works. 
Since $\tilde\BQ_\DF$ and $\BI + \tilde\BQ_\DF$ commute, 
they share a common set of eigenvectors and they can be 
expressed in the same reduced basis.
Therefore, these expressions hold for the compressed whitened data presented in
\citet{vanhaasterenCalculationPvaluesQuadratic2025}. 
This makes it straightforward to
compute the $\NPMV$ filter and $p$-values with minimal modifications to existing
codes.

\section{Derivation of the \texorpdfstring{$\NPMV$}{NPMV} filter}
\label{s:NPW-derivation}

  In this section, we derive the form of the $\NPMV$ filter defined
  by~\eqref{e:argmin}.  For notational simplicity, we will assume in
  this section that the data $z$ have only a single sample per pulsar,
  $z=\{z_a\}$, where $a=1,2,\cdots,n$ runs over the number of pulsars
  in the array.  Nonetheless, the results hold for the more general
  multi-sample case.

  For most of this section, $\BN$ is an arbitrary Hermitian,
  positive-definite matrix. However, at the end, we restrict attention
  to block-diagonal $\BN$ as defined by~\eqref{e:blockdiag}.  This
  assumption is needed to obtain the simple form of $\BQ_\NPMV$ given in
  \eqref{e:npw_filter}.

The $\NPMV$ detection statistic is defined by~\eqref{e:argmin}:
$D_\NPMV$ is the quadratic form whose difference with 
$D_\NP$ has the smallest possible variance under $\HN$, 
subject to the constraint that the filter $\BQ_\NPMV$ does 
not use autocorrelations.  
Thus, $\BQ_\NPMV$ minimizes
\be
V(\BQ) \equiv {\rm Var}[D(z|\BQ)-D(z|\BQ_\NP)]_{\HN}
\label{e:min_diff2}\\
\ee
subject to the constraint
\be \Diag Q = 0 \iff Q_{aa} = 0\,,
\label{e:constraint}
\ee
where the pulsar index $a$ is not summed.
To simplify the expressions that follow, we define
\be
\BDelta \equiv \BQ - \BQ_\NP\,,
\label{e:Delta}
\ee
for which 
\be
V(\BQ) = {\rm Var}[D(z|\BDelta)]_{\HN}\,.
\label{e:VDelta}
\ee
Expanding the variance of $\BDelta$ in terms of the 
data $z$, we find:
\be
\begin{aligned}
V(\BQ)
&=\left\langle |D(z\,|\,\BDelta)|^2\right\rangle_{\HN}
- |\langle D(z\,|\,\BDelta)\rangle_{\HN}|^2
\\
&=\sum_{a,b,c,d}\Delta_{ab}\Delta_{cd}^* 
\left(\left\langle z_a^* z_b z_c z_d^*\right\rangle_{\HN}
-\left\langle z_a^* z_b\right\rangle_{\HN}\left\langle z_c z_d^*\right\rangle_{\HN}\right)
\\
&=\sum_{a,b,c,d}\Delta_{ab}\Delta_{cd}^* 
\left\langle z_a^* z_c\right\rangle_{\HN}\left\langle z_b z_d^*\right\rangle_{\HN}
\\
&=\sum_{a,b,c,d}\Delta_{ab}\Delta_{cd}^* N^*_{ac} N_{bd}
={\rm Tr}(\BDelta\BN\BDelta\BN)\,,
\label{e:obj2}
\end{aligned}
\ee
where the third equality follows from 
Isserlis's theorem~\citep{isserlisFormulaProductmomentCoefficient1918},
the fourth equality from $\langle z_a z_b^*\rangle_{\HN}\equiv N_{ab}$, and the
fifth from the Hermitian property of $\BN$ and $\BDelta$.

We now use the method of Lagrange multipliers to solve the
constrained optimization problem.
We optimize
\be
L(\Lambda,\BQ)
\equiv{\rm Tr}(\BDelta\BN\BDelta\BN)- \sum_a \Lambda_a Q_{aa}\,,
\ee
where $\Lambda_a$ are the Lagrange multipliers (one per diagonal element
of $\BQ$).
Setting the variations of $L$ wrt $\Lambda_a$ 
and $Q_{ab}$ equal to zero gives
\begin{align}
&\delta\Lambda_a:\quad Q_{aa} = 0\,,
\label{e:dLambda}\\
&\delta Q_{ab}:\quad
2\sum_{c,d} N_{bd}\Delta_{dc}N_{ca} -\Lambda_a\,\delta_{ab} =0\,.
\label{e:dQab}
\end{align}
The first equation is the constraint \eqref{e:constraint}. 
To work with the second equation, we multiply it by the inverse 
matrices $(N^{-1})_{eb}$ on the left and $(N^{-1})_{af}$ 
on the right; the matrix multiplication sums
over all values of $a$ and $b$.  
The result can be rewritten as
\be
\Delta_{ef} = \frac{1}{2}\sum_a (N^{-1})_{ea} \Lambda_a (N^{-1})_{af}\,,
\ee
or, equivalently,
\be
Q_{ab} = Q^\NP_{ab}+\frac{1}{2}\sum_c (N^{-1})_{ac} \Lambda_c (N^{-1})_{cb}\,,
\label{e:soln_Qab}
\ee
where we have relabeled the indices in the last equation and used 
\eqref{e:Delta} to express $\BDelta$ in terms of $\BQ$ and $\BQ_\NP$.

The equations for the desired filter $\BQ_\NPMV = Q_{ab}$ have a unique
solution. To see this, consider the diagonal and off-diagonal
components of \eqref{e:soln_Qab}.  These can be written
\begin{align}
&0= Q^\NP_{aa} + \frac{1}{2}\sum_c (N^{-1})_{ac} \Lambda_c (N^{-1})_{ca}\,,
\label{e:soln_Qaa}
\\
&Q_{\bar a\bar b}
=Q^\NP_{\bar a\bar b} 
+\frac{1}{2}
\sum_c (N^{-1})_{\bar a c} \Lambda_c (N^{-1})_{c\bar b}\,,
\label{e:soln_Qabarbbar}
\end{align}
where barred indices range over $1,2,\cdots,n$, but must not equal one
another.  The diagonal equations \eqref{e:soln_Qaa} arise from the
constraint $Q_{aa}=0$.  They relate the Lagrange multipliers
$\Lambda_a$ to the diagonal components of $\BQ_\NP$ and the noise
covariance matrix $\BN$, and are $n$ linear equations in the $n$
variables $\Lambda_a$.  Since the linear transformation
\be
F_{ac} \equiv (N^{-1})_{ac} (N^{-1})_{ca} = |(N^{-1})_{ac}|^2
\ee
is invertible%
\footnote{Since $\BN$ is Hermitian and positive-definite, this follows
from the Schur product theorem (see Assertion (c) of (7.5.3) in \cite{Horn-Johnson:2012}, 
 with $\BA=\BN^{-1}$ and
$\BB=(\BN^{-1})^*$).} they provide a unique solution for
$\Lambda_a$.  Substituting these $\Lambda_a$ into the rhs of
\eqref{e:soln_Qabarbbar} provides the unique solution for $Q_{\bar a\bar b}$. 
$\BQ_\NPMV$ then follows immediately, since from~\eqref{e:dLambda} 
the remaining (diagonal) terms vanish.

This filter $\BQ$ does not have the simple form given in
\eqref{e:npw_filter}.
To obtain \eqref{e:npw_filter}, we need to additionally 
assume that $\BN$ is block-diagonal, which is a 
standard assumption for noise covariance matrices; see 
Sec.~\ref{s:CURN}.
Then $N_{ac}= N_a\,\delta_{ac}$, implying
\be
F_{ac} = N_a^{-2}\, \delta_{ac}\,.
\ee
This immediately leads to
\be
\Lambda_a = -2 N_a^2 Q_{aa}^\NP
\label{e:Lambda}
\ee
as the solution to~\eqref{e:soln_Qaa}.  It also implies that the
second term on the rhs of~\eqref{e:soln_Qabarbbar} vanishes:
\be
\sum_c (N^{-1})_{\bar a c} \Lambda_c (N^{-1})_{c\bar b}
= N_{\bar a}^{-2} \Lambda_{\bar a} \delta_{\bar a\bar b} = 0\,.
\label{e:zero}
\ee
The final equality follows from 
$\delta_{\bar a\bar b}=0$ for $\bar a\ne\bar b$, 
independent of the solution for $\Lambda_a$.
Thus, for block-diagonal $\BN$, we see from \eqref{e:zero} 
and \eqref{e:soln_Qabarbbar} that 
\be
Q_{\bar a\bar b} = Q^\NP_{\bar a\bar b}\,.
\label{e:Qoffdiag}
\ee
Then \eqref{e:dLambda} and \eqref{e:Qoffdiag} 
give the full solution
\be
Q^\NPMV_{aa}=0\,,\quad
Q^\NPMV_{\bar a\bar b} = Q^\NP_{\bar a\bar b}
\quad(\bar a\ne\bar b)\,,
\ee
which may also be written as \eqref{e:npw_filter}.

\section{Generalized \texorpdfstring{$\chi^2$}{ChiSquared} distribution}
\label{s:GX2}

\substantialchange{For a multivariate normal distribution, the quadratic statistic
$D(z\,|\,\BQ)$ in~\eqref{e:quad_DS} follows a generalized $\chi^2$
distribution. Hence the false-alarm and detection probabilities for a
threshold $\tau$ are given by the corresponding complementary
cumulative distribution function (CDF) under $\HN$ and $\HS$. A
discussion of these CDFs, and their application to PTAs (for example,
computing $p$-values) may be found in~\cite{Hazboun-et-al:2023}.  This
also contains citations to the more general literature on this topic,
for example~\cite{dasMethodIntegrateClassify2021}.}

The general case has a real data space, a normal distribution defined
by a real positive-definite symmetric covariance matrix, and a
statistic~\eqref{e:quad_DS} defined by a real symmetric matrix
(filter) $\BQ$.  For this general case, there is no known closed form
for the CDF.  However, there are many different algorithms and computer
packages to evaluate it~\citep{dasMethodIntegrateClassify2021}.

The complex normal distribution of dimension $n$ is a special case.
It is equivalent to a real data space of dimension $2n$, where (in a
suitable basis) the real random variables come in pairs $(x_j,y_j)$,
where $x_j$ and $y_j$ are statistically independent of one another and
have identical variance for $j=1,\dots,n$. Then, the distribution of
$x_j^2 + y_j^2$ is exponential, and the CDF factors into a product of
exponentials.  For this special case, an explicit closed form for the
CDF is obtained~\citep[pg.~17~Eq.~(4)]{cox_renewal_1962} using Laplace
transforms.  Those results assume that the statistic $D(z|\BQ)$ given
in \eqref{e:quad_DS} is positive or negative-definite.

Here, we are also interested in matrices (filters) $\BQ$ which vanish
on the diagonal. Since $\Tr \BQ = 0$, it follows that $\BQ$ has
eigenvalues of both signs.  With this as motivation, we provide a
derivation of the CDF for the indefinite case, which we use to
characterize and optimize decision rules.

\subsection{Definitions}

Let $z$ denote a complex column vector, $\BC$ a Hermitian
positive-definite covariance matrix, and $\BQ$ a Hermitian matrix, all
of dimension $n$.  We assume that $\BQ$ has no vanishing eigenvalues:
it is full rank.  The cases of interest are $\BC \to \BN$ for the
null distribution and $\BC \to \BS$ for the signal distribution.

The CDF for a generalized (central) $\chi^2$ distribution with $n$
complex degrees of freedom [as given in \eqref{e:data}, for example]
is
\begin{equation}
  \label{e:def}
  F(\tau,\BC,\BQ) 
\equiv \frac{1}{\det (\pi \BC)} \int \lb \dd^n z \,
  \theta(\tau - z^\dagger \BQ z) \, {\rm e}^{- z^\dagger \BC^{-1} z} \, ,
\end{equation}
with the Heaviside step function $\theta(x)$ defined by
\eqref{e:stepfn}.  Here, $\dd^n z \equiv \prod_{j=1}^n \dd\,{\rm
  Re}(z_j) \, \dd\,{\rm Im}(z_j)$, and the normalization is such that
$\langle z z^\dagger \rangle = \BC$, consistent with the convention
$\langle z_a z_b^* \rangle = N_{ab}$ used in
Sec.~\ref{s:NPW-derivation}.  $F$ is the probability that the quadratic form $D=
z^\dagger \BQ z$ is less than a real threshold $\tau$; $F$ vanishes as
$\tau \to -\infty$, and approaches $1$ as $\tau \to +\infty$.  For
$\BQ$ positive (negative) definite, $F$ has its support on the
positive (negative) $\tau$-axis. If $\BQ$ is indefinite, then $F$ is
nonzero for all $\tau$.  \substantialchange{The function $\barF(\tau,\BC,\BQ) \equiv
1-F(\tau,\BC,\BQ) = F(-\tau,\BC,-\BQ)$ is called the \emph{survival
function} or \emph{complementary CDF}.}

In the space of possible data $z \in \mathbb{C}^{n}$, the normal
distribution $\propto \exp(-z^\dagger \BC^{-1} z)$ has level
surfaces which are ellipsoids, with principal-axis directions and
squared lengths given by the eigenvectors and eigenvalues of $\BC$.
The statistic $D=z^\dagger \BQ z$ has level surfaces which are
ellipsoids if $\BQ$ is positive or negative definite.  If $\BQ$ has
indefinite signature $(-,\dots,-,+,\dots,+)$, then the level surfaces
are nested hyperbolic sheets.

\substantialchange{The false alarm and false dismissal probabilities may be written in
terms of $F$ and $1-F$.  In the notation of~\eqref{e:<>},
$F(\tau,\BN,\BQ) = \langle \theta(\tau - z^\dagger \BQ z)
\rangle_{H_N} = 1 - \langle \theta(D(z|\BQ)-\tau) \rangle_{H_N}$. So
from \eqref{e:FAP} it follows that the false alarm probability $\alpha
= 1 - F(\tau, \BN, \BQ) = \barF(\tau,\BN,\BQ) = F(-\tau, \BN,
-\BQ)$. In a similar vein, from \eqref{e:DP} the detection probability
is $1-\beta = 1 - F(\tau, \BS, \BQ) = \barF(\tau,\BS, \BQ)= F(-\tau,
\BS, -\BQ)$, and the false dismissal probability is $\beta = F(\tau,
\BS, \BQ)$.}

\subsection{Evaluation of \texorpdfstring{$F$}{CDF}}

To evaluate $F\equiv F(\tau,\BC,\BQ)$, 
we first represent the Heaviside step function as the
Fourier integral
\begin{equation}
  \label{e:theta1}
  \theta(x) = \lim_{\epsilon \to 0^+}\int_{-\infty}^\infty \bs \dd f \, \frac{{\rm e}^{2 \pi i f x}}{2 \pi i (f - i \epsilon)} \, ,
\end{equation}
where $\epsilon > 0$ is real.  The derivative of \eqref{e:theta1} with
respect to $x$ yields the standard expression for the Dirac delta
function.

\begin{figure}[t]
\centering
\includegraphics[width=0.98\linewidth]{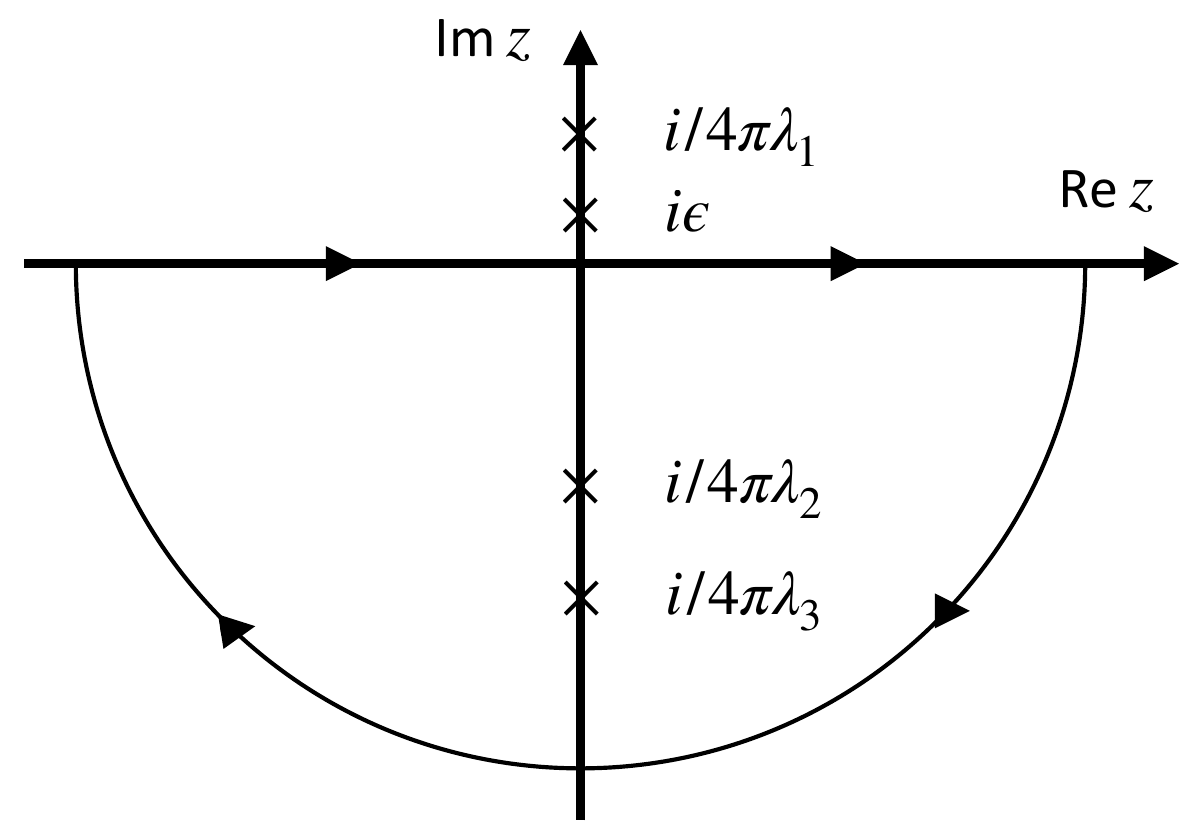}
\caption{In the complex $z$-plane, the integrand of \eqref{e:F3} is
  analytic apart from $n+1$ poles along the imaginary axis.  We
  illustrate the 3-dimensional case for which $\BE$ has one positive
  and two negative eigenvalues.  If $\tau<0$, the integral along the
  real axis is evaluated by closing the contour of integration in the
  lower half-plane, and using Cauchy's residue theorem.}
\label{fig:ComplexPlane}
\end{figure}

One can verify \eqref{e:theta1} by inspection, using Cauchy's residue
theorem.  For $x < 0$, the exponential ${\rm e}^{2 \pi i f x}$ falls
off exponentially as $f \to -i \infty$, so one can close the path of
integration with a clockwise half circle in the lower half complex
$f$-plane, similar to that shown in Fig.~\ref{fig:ComplexPlane}.  Since the pole at
$f = i \epsilon$ lies above the real axis, the contour of integration
does not contain any poles, so the sum of the residues vanishes and
$\theta(x) = 0$.  For $x > 0$ one can close the path of integration
with a counterclockwise half circle in the upper half complex
$f$-plane.  The residue of the pole is $1/2 \pi i$, so the residue
theorem gives $\theta(x) = 1$.  For $\theta(0)$ take the principle
value.

Substituting \eqref{e:theta1} into \eqref{e:def} gives
\begin{equation}
  \label{e:F1}
  F  \! = \! \lim_{\epsilon \to 0^+}\int_{-\infty}^\infty \! \bs \dd f \,
  \frac{{\rm e}^{2 \pi i f \tau}}{2 \pi i (f - i \epsilon)}
  \! \int \lb \dd^n z  \,\frac{{\rm e}^{- z^\dagger \bigl[ \BC^{-1}  + \, 2 \pi i f \BQ \bigr] z}}{\det(\pi\BC)} \, .
\end{equation}
The integral over $z$ is easily evaluated, giving
\begin{eqnarray}
  \nonumber
  F & = & \lim_{\epsilon \to 0^+}\int_{-\infty}^\infty \! \bs \dd f \,
  \frac{{\rm e}^{2 \pi i f \tau}}{2 \pi i (f - i \epsilon)} 
  \frac{\det \bigl(\pi \bigl[ \BC^{-1}  + 2 \pi i f \BQ \bigr]^{-1}\bigr)}{\det(\pi \BC)}  \\
  \label{e:F2}
  & = &\lim_{\epsilon \to 0^+}\int_{-\infty}^\infty \! \bs \dd f \,
  \frac{{\rm e}^{2 \pi i f \tau}}{2 \pi i (f - i \epsilon)} 
  \frac{1}{\det \bigl( \BI  + 2 \pi i f \BE \bigr)} \, ,
\end{eqnarray}
where
\be
\BE \equiv \BC^{1/2} \, \BQ \, \BC^{1/2} \, .
\label{e:DefE}
\ee
Here, $\BC^{1/2}$ denotes the unique Hermitian, positive-definite
matrix satisfying $\BC = \BC^{1/2} \BC^{1/2}$.

Let $\lambda_j$ for $j=1,2, \dots, n$ denote the (real) eigenvalues of
the Hermitian matrix $\BE$, which we assume are
distinct (non-degenerate).  Then, since $\BI$ commutes with $\BE$,
\begin{equation}
    \begin{aligned}
  \label{e:F3}
    F& = \lim_{\epsilon \to 0^+}\int_{-\infty}^\infty \bs \dd z \,
    \frac{{\rm e}^{2 \pi i \tau z}}
    {2 \pi i (z - i \epsilon) \prod_j ( 1  + 2 \pi i \lambda_j z )} \\
    & = \frac{1}{\det( 2 \pi i \BE)} \lim_{\epsilon \to 0^+}\int_{-\infty}^\infty \bs \dd z \,
    \frac{{\rm e}^{2 \pi i \tau z}}{2 \pi i (z - i \epsilon) \prod_j ( z  - z_j)} \, .
      \end{aligned}
\end{equation}
Here, we changed the name of the integration variable $f$ to the
complex variable $z=x+iy$ [not to be confused with the complex data
vector $z$, which was integrated away in~\eqref{e:F1} to obtain~\eqref{e:F2}]. We also defined
\begin{equation}
  \label{e:zj}
  z_j \equiv \frac{i}{2 \pi \lambda_j} \, .
\end{equation}
These are a set of $n$ purely imaginary values determined by the eigenvalues of $\BE$.

The integrand in \eq{e:F3} is a meromorphic function on the complex
$z$-plane, which is holomorphic except for isolated simple poles at
$z=z_j$ and $z = i \epsilon$ along the imaginary axis, as illustrated
in Fig.~\ref{fig:ComplexPlane}.  The numerator is analytic everywhere,
and the denominator grows $\propto R^{n+1}$ for large $R \equiv
|z|$. Hence, one can apply arguments identical to those following
\eqref{e:theta1}.  For $\tau < 0$ we close the path of integration
with a half-circle in the lower half $z$-plane which encloses all of
the poles along the negative imaginary axis, as shown in
Fig.~\ref{fig:ComplexPlane}. The sum of their residues gives the
integral via Cauchy's theorem.

\substantialchange{To simplify matters, first assume that one eigenvalue $\lambda_1 < 0$
of $\BE$ is negative, that the remaining eigenvalues are positive,
that all of the eigenvalues are distinct, and that $\tau < 0$.  Then,
the integration contour encloses a single pole located at $z = z_1$, and
Cauchy's residue theorem gives
\begin{equation}
  \label{e:F4}
   F(\tau, \BC, \BQ) = 
     \frac{-2 \pi i}{\det( 2 \pi i \BE)} 
    \frac{{\rm e}^{2 \pi i \tau z_1}}{2 \pi i z_1 \prod_{j \ne 1} ( z_1  - z_j)} \, ,
\end{equation}
where the minus sign is because the contour of integration is clockwise.
Substituting \eq{e:zj} and canceling factors gives
\begin{equation}
  \label{e:F5}
   F(\tau, \BC, \BQ) 
   = {\rm e}^{- \tau/\lambda_1} \prod_{j \ne 1} \left( \frac{\lambda_1}{\lambda_1 - \lambda_j} \right) \, \text{ for } \tau < 0 \, .
\end{equation}
If there are multiple negative eigenvalues, then this formula must be
summed over them.

We easily obtain $F$ for $\tau > 0$.  Since
$\theta(x) = 1 - \theta(-x)$, it follows immediately from \eq{e:def}
that $F(\tau, \BC, \BQ) = 1 - F(-\tau, \BC, -\BQ)$, because
changing the sign of $\BQ$ changes the signs of the eigenvalues of $\BE$.

Thus, the full formula for $F$ may be written in terms of the $n$
(distinct and nonzero) eigenvalues $\lambda_j$ of $\BE \equiv \BC^{1/2}
\, \BQ \, \BC^{1/2}$ as
\begin{equation}
\label{e:F6F7}
\mspace{-20mu}F(\tau, \BC, \BQ)=
\begin{cases}
\begin{aligned}
\phantom{1} \phantom{-} \!\!\!&\sum_{j \mid \lambda_j < 0} e^{-\tau/\lambda_j}
\prod_{k \ne j} \left(\frac{\lambda_j}{\lambda_j-\lambda_k} \right)
\end{aligned}
& \mspace{-25mu} \text{if }  \tau \le 0,\\[1.2ex]
\begin{aligned}
1-&\!\!\!\sum_{j \mid \lambda_j > 0} e^{-\tau/\lambda_j}
\prod_{k \ne j} \left(\frac{\lambda_j}{\lambda_j - \lambda_k} \right)
\end{aligned}
& \mspace{-25mu} \text{if }  \tau \ge 0 .
\end{cases}
\end{equation}
Note that a special case of \eqref{e:F6F7}, where all eigenvalues have
the same sign, is derived in~\citep[pg.~17~Eq.~(4)]{cox_renewal_1962}
using Laplace transforms; \eqref{e:F6F7} is a generalization to the
indefinite case.  We subsequently found this in the information theory
literature~\cite{Earlier2Fformula}.  One can prove that the upper and
lower expressions in \eqref{e:F6F7} are equal as $\tau \to 0$ from
either side (also see remarks at the end of Sec.~\ref{ss:numeval}).

Notes: (a) The products appearing in \eq{e:F6F7} include $n-1$ terms.
(b) If there are no positive eigenvalues, then $F=1$ for $\tau \ge 0$.
(c) If there are no negative eigenvalues, then $F= 0$ for $\tau \le
0$. (d) The survival function (complementary CDF) is $\barF(\tau,
\BC, \BQ) \equiv 1 - F(\tau, \BC, \BQ) = F(-\tau, \BC, -\BQ)$.}

\substantialchange{
\subsection{Numerical evaluation of the CDF \texorpdfstring{$F(\tau,\BC,\BQ)$}{FCDF}}
\label{ss:numeval}

We now discuss the numerical evaluation of \eqref{e:F6F7}, which,
given the $n$ eigenvalues $\lambda_j$ of $\BE \equiv \BC^{1/2} \, \BQ \,
\BC^{1/2}$, appears straightforward. In practice, however, it is not:
the calculation is complicated by a well-known numerical issue,
\emph{catastrophic cancellation}.

It is easy to see why this issue arises. We write \eq{e:F6F7} as
\begin{equation}
  \label{e:fexp}
F(\tau, \BC, \BQ)=
\begin{cases}
\begin{aligned}
\phantom{1} \phantom{-} \!\!\!&\sum_{j \mid \lambda_j < 0} c_j \, e^{-\tau/\lambda_j}
\end{aligned}
&  \text{for }  \tau \le 0,\\[1.5ex]
\begin{aligned}
1-&\!\!\!\sum_{j \mid \lambda_j > 0} c_j \, e^{-\tau/\lambda_j}
\end{aligned}
& \text{for }  \tau \ge 0 ,
\end{cases}
\end{equation}
where the coefficients $c_j$ of the exponential factors are
\begin{equation}
\label{e:bigfactor}
c_j \equiv \prod_{k \ne j} \left(\frac{\lambda_j}{\lambda_j - \lambda_k} \right) \, .
\end{equation}
Because of the constraints on the signs of $\lambda_j$ and $\tau$, the
exponential factors that appears in \eq{e:fexp} lie in the range
$\exp(-\tau/\lambda_j) \in (0,1]$.  However, the factors $c_j$ that
  multiply them can have either sign, and in practice are \emph{very}
  large compared to one.

In our single-frequency-bin example, the $\NPMV$ filter for the $n=67$
NG pulsars has $16$ positive eigenvalues and $51$ negative
eigenvalues; the largest magnitude coefficients have $|c_j| \approx
10^{32}$.  For $n=121$ pulsars analyzed in \cite{FivePTA}, the largest
magnitude coefficients for $\NPMV$ have $|c_j| \approx 10^{60}$.  For
$150, 200, 250$, and $300$ pulsars randomly located on the sky, the
largest magnitude coefficients for $\NPMV$ have $|c_j| \approx
10^{103}, 10^{141}, 10^{185}$ and $10^{226}$.

The numerical challenges arises because the CDF lies in the range
$F(\tau, \BC, \BQ) \in [0, 1]$, and we typically need its value to
high accuracy.  For example, a $5\sigma$ FAP corresponds to $1-F
\approx 2.87 \times 10^{-7}$. Suppose that we want to compute $F$ with
errors $\epsilon < 10^{-8}$.  This means that the sum of the large
$c_j$ (multiplied by factors in the range $(0,1]$ from the
  exponentials) must be accurate 8 decimal digits to the right of the
  decimal point, but the individual terms being added can have several
  hundred decimal digits to the left of the decimal point.  IEEE 754
  double precision floating point numbers have one sign bit, a 52-bit
  mantissa, and 11 exponent bits, corresponding to $52\log_{10}(2)
  \approx 15.65$ decimal digits of resolution~\citep[][Chapter~1,
    Sec.~1.3]{NumericalRecipes}. This is far less than needed for the
  worst-case example above; to get the desired resolution requires
  arbitrary precision math libraries capable of computing accurately
  with $\approx 800$ bit mantissa.

In the supplementary materials for this paper, we provide such an
implementation, using the \texttt{mpmath} arbitrary precision
floating-point math library~\citep{mpmath}. (Note that \texttt{mpmath}
is substantially faster if used with the \texttt{gmpy2}
library~\citep{gmpy2} backend.)  Our implementation first uses standard
IEEE 754 double-precision to compute
\begin{align}
  \label{e:bits}
b_L &= \max\left(\max_{j|\lambda_j <  0} \left( -\tau/\lambda_j + \log | c_j|\right),0\right)/\log(2) \\
b_R &= \log_2\!\left(\frac{\sqrt{m}}{\epsilon}\right) \, ,
\end{align}
in floating point, if $\tau \le 0$. [The sign of the eigenvalues in the summation of \eqref{e:bits} is
  reversed if $\tau > 0$.]  To obtain $\log|c_j|$ we sum the logs of
the $n-1$ factors in $c_j$. After maximizing over $j$, the resulting
$b_L$ is the number of bits to the left of the binary point which are
needed to represent all of the $c_j \exp(-\tau/\lambda_j)$ with an
error of order unity.  An additional $b_R$ bits are required to the
right of the binary point to obtain the desired precision
$\epsilon$. [Here, $m$ is the number of terms in the
  sum~\eqref{e:fexp} and we assume that summation/multiplication
  errors add in quadrature. In practice the multiplication errors
  appear to be negligible.] Hence, the total required number of
mantissa bits is $b = { \rm ceil} (b_L + b_R)$.  If $b < 53$, then we
complete the computation of~\eqref{e:fexp} using IEEE 754
double-precision floats.  If $b \ge 53$, then we compute~\eq{e:F6F7}
using arbitrary precision floating point operations, after setting the
internal precision to $b$ mantissa bits.

We tested this implementation using from $30$ to $300$ pulsar
locations, and using $\BQ$ filters for the $\NP$, $\NPMV$ and $\DF$
statistics.  For all of these, we computed the CDF for both the null
($\BC \to \BN$) and signal ($\BC \to \BS$) cases, over a wide range of
thresholds corresponding to FAP larger than $10^{-8}$. The worst-case
number of mantissa bits is approximately five times the number of
pulsars minus 76. In testing, the average computation time scales
approximately quadratically with the number of pulsars. With 300
pulsars the mean run time was about 10 milliseconds on a Mac M4
laptop. Note that run time depends strongly upon $\tau$ and seems to
be longer for negative $\tau$ than for positive $\tau$.

It may be possible to accurately compute $F$ using only standard
double-precision arithmetic.  One can compute $F(\tau=0, \BC, \BQ)$ in
this way, using convex interpolation.  This is done by introducing the
Lagrange polynomials
\begin{equation}
\ell_j(x) = \prod_{k \ne j} \frac{x - \mu_k}{\mu_j - \mu_k} \, ,
\end{equation}
where the weights $\mu_j = 1/\lambda_j$, ordered so that $\mu_1 <
\cdots < \mu_m < 0 < \mu_{m+1} < \cdots < \mu_n$.  The $\ell_j(x)$ are
polynomials of order $n-1$ which satisfy $\ell_j(\mu_k) = \delta_{jk}$
and $c_j = \ell_j(0)$. Thus, $P(x) = \ell_1(x) + \dots + \ell_m(x)$ is
a polynomial in $x$ of order $n-1$, which has value unity for
$x=\mu_1, \dots , \mu_m$ and vanishes for $x=\mu_{m+1}, \dots ,
\mu_n$. By counting extrema, it is easy to see that $P(x)$ is
monotonic between $P(\mu_m) = 1$ and $P(\mu_{m+1}) = 0$, and since
$F(\tau=0) = P(0)$, this demonstrates that $F(\tau=0)$ lies in the
interval $[0,1]$. Moreover, Neville’s algorithm provides a convex
interpolation to obtain $F(\tau=0) \in [0,1]$ with ordinary floating
point.  However, we have not found a way to extend this to $\tau \ne
0$.
}

\section{Statistic Performance Comparison}
\label{s:comparison}

Here, we use ROC curves to compare the performance of four different
quadratic statistics of the form \eqref{e:quad_DS}. The first is
$\NP$, which (by construction) has the best performance.  Next is the
(numerically-constructed) $\NPCC$ filter, which maximizes the detection
probability at fixed FAP, subject to the constraint that it uses only
cross-correlations between pulsars. Third is the $\NPMV$ statistic, 
which is the quadratic statistic that
is ``as close as possible" to the Neyman-Pearson-optimal statistic,
also under the constraint that it uses only cross-correlations, 
see \eqref{e:argmin}.  In last place is the $\DFCC$ or
$\DF$ statistic, which are identical for the CURN null hypothesis
that we consider for our models.

We employ a toy model for the data.  This enables the $\NPCC$ filter
to be found by numerically optimizing the analytical form of $F$, as
obtained in Sec.~\ref{s:GX2} for a generalized $\chi^2$ distribution.
(Note that for realistic data we cannot use the analytical
form~\eqref{e:F6F7}, because the data cannot be
described in terms of a complex-valued random process.)

\subsection{Toy Model}
\label{ss:ToyModel}

Our synthetic PTA contains $n=67$ pulsars whose sky locations are chosen to
match the NANOGrav 15-year data set~\citep{agazieNANOGrav15Yr2023b}.
The direction to each pulsar on the sky is
represented by a unit vector $\hat{n}_a$, where the index $a = 1,
\dots, n$ labels the pulsars.  For each pulsar, the data consists of a
single complex number $z_a$.  The full dataset is represented as an
$n$-dimensional column vector $z$, whose components are the individual
$z_a$.

For the signal hypothesis $\HS$, the $z$ are described by a zero-mean
complex normal distribution
\begin{equation}
  \label{e:H0toy}
  z \sim \CN(0, \BS) \, ,
\end{equation}
with a real covariance matrix $\langle z z^\dagger \rangle = \BS$,
where
\begin{equation}
  \BS \equiv \ch^2\mu_{ab} + \sigma^{2}\delta_{ab} \, .
  \label{e:data}
\end{equation}
The pulsar noise variance is $\sigma^2 = 1$ and the GWB variance is 
$\ch^2 = 1$.
The  HD matrix is
\begin{equation}
  \begin{aligned}
    \label{e:HDmat}
    \mu_{ab} \equiv  \, & \frac{1}{2}(1 + \delta_{ab}) - \frac{1}{8}(1 - \hat n_a \cdot \hat n_b) \, +  \\
    & \frac{3}{4}(1 - \hat n_a \cdot \hat n_b) \log \left( \frac{1 - \hat n_a \cdot \hat n_b}{2} \right) \, .
  \end{aligned}
\end{equation}
Note that this is often written in terms of $\cos \gamma_{ab} = \hat n_a \cdot
\hat n_b$, where $\gamma_{ab}$ is the angle between the lines of sight to
pulsars $a$ and $b$.

For the null hypothesis $\HN$, we use CURN as discussed in
Sec.~\ref{s:CURN}.  The $\HN$ data are described by $z \sim \CN(0,
\Diag \BS)$, so the noise covariance is $\BN = \Diag \BS$ as given
in~\eqref{e:CURN}.  This CURN hypothesis preserves the GWB induced
autocorrelations, but eliminates the HD cross-correlations. It is
considered ``more conservative'' than the alternative, which would be
to set $\ch = 0$ in~\eqref{e:data} giving 
$\BN \equiv N_{ab} = \sigma^2\delta_{ab}$.

\subsection{ROC curves for Toy Model}
\label{ss:ROC_toy}

For the toy model, we form the $\DFCC$, $\NP$, and $\NPMV$ filters as
in \eqref{e:defcc_filter}, \eqref{e:ump_filter}, and
\eqref{e:npw_filter}.  To construct the $\NPCC$ filter, we solve the
full numerical optimization problem, maximizing the detection
probability at fixed FAP, using the exact expressions for the CDF
in~\eqref{e:F6F7}.  This is a demanding task: the
objective function is high-dimensional and strongly multi-modal, with
a large region of parameter space whose attraction basins pull the
optimization towards the $\DFCC$ filter.  In practice, we find that
from generic initial conditions, many optimizers converge to $\DFCC$
rather than to the desired cross-correlation $\NPCC$ solution.

To overcome this unwanted convergence to $\DFCC$, we employ an
ensemble search strategy, combining the stochastic {\it simultaneous
  perturbation stochastic approximation} (SPSA)
method~\citep{Spall1998SPSA}---whose stochastic updates resemble an
annealing process---with subsequent local refinement using the
derivative-free BOBYQA algorithm~\citep{Powell2009BOBYQA}.  By
launching many such searches from diverse starting points, and in
particular from the vicinity of the $\NPMV$ filter, we consistently
find solutions whose ROC curves lie above those of $\NPMV$.  This
demonstrates that the true $\NPCC$ filter is close to $\NPMV$ in form
but achieves higher detection probability.  Although we cannot
guarantee convergence to the unique global maximum, the consistency of
our toy model results across independent ensemble runs indicates that we are
closely approaching the genuine $\NPCC$ solution.

\begin{figure*}[t] 
\centering
\includegraphics[width=0.49\linewidth]{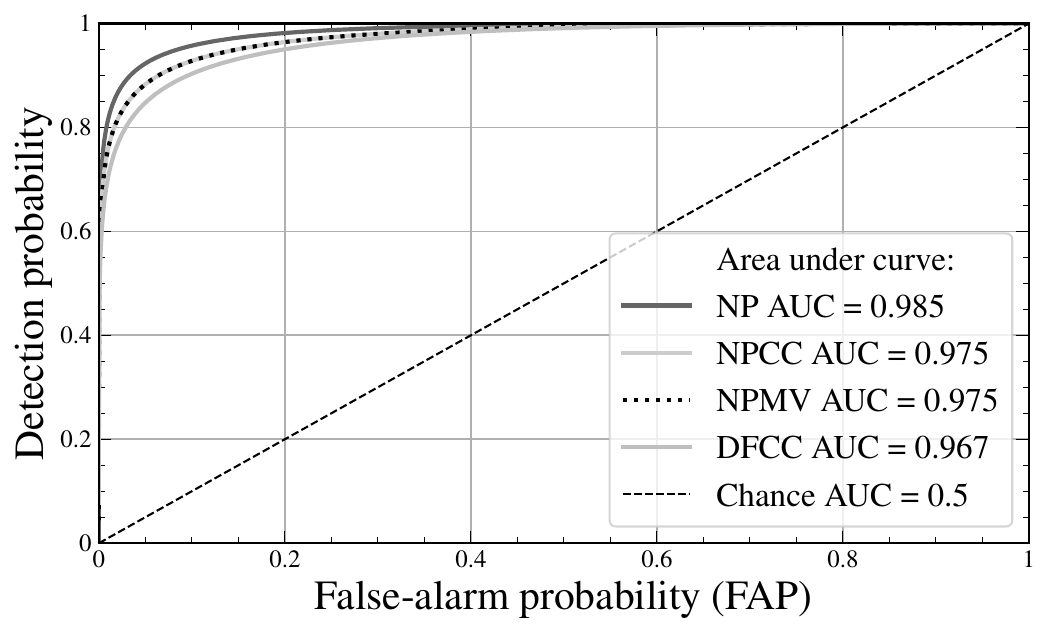}
\includegraphics[width=0.49\linewidth]{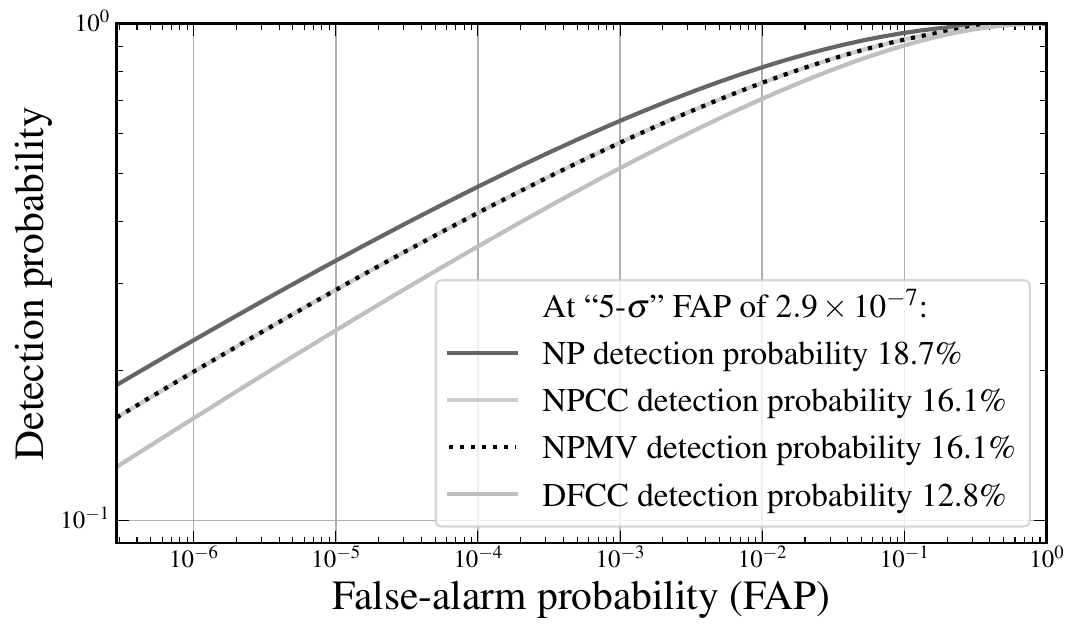}
\caption{ROC curves of different detection statistics, for the
  toy model of Sec.~\ref{ss:ToyModel}.
  These compare the standard
  ``optimal" cross-correlation deflection statistic $\DFCC$, 
  the Neyman-Pearson-optimal statistic $\NP$, 
  the cross-correlation-only Neyman-Pearson statistic $\NPCC$, and
  the cross-correlation-only Neyman-Pearson-Minimum-Variance statistic $\NPMV$.
  Left panel: linear scale.
  Legend gives AUC ``area under the ROC curve" for the 
  different statistics (larger is better).
  Right panel: same as left panel, but for logarithmic axes.
  Legend gives detection probability for the different
  statistics at the $5$-$\sigma$ discovery 
  threshold FAP of $2.9 \times 10^{-7}$ (leftmost
  plot boundary).}
\label{fig:ROC_TOY}
\end{figure*}

The ROC curves of the various different detection statistics are shown in
Fig.~\ref{fig:ROC_TOY}.

\subsection{ROC curves for more realistic model}
\label{ss:ROCcurves}

\begin{figure}[t]
\centering
\includegraphics[width=0.98\linewidth]{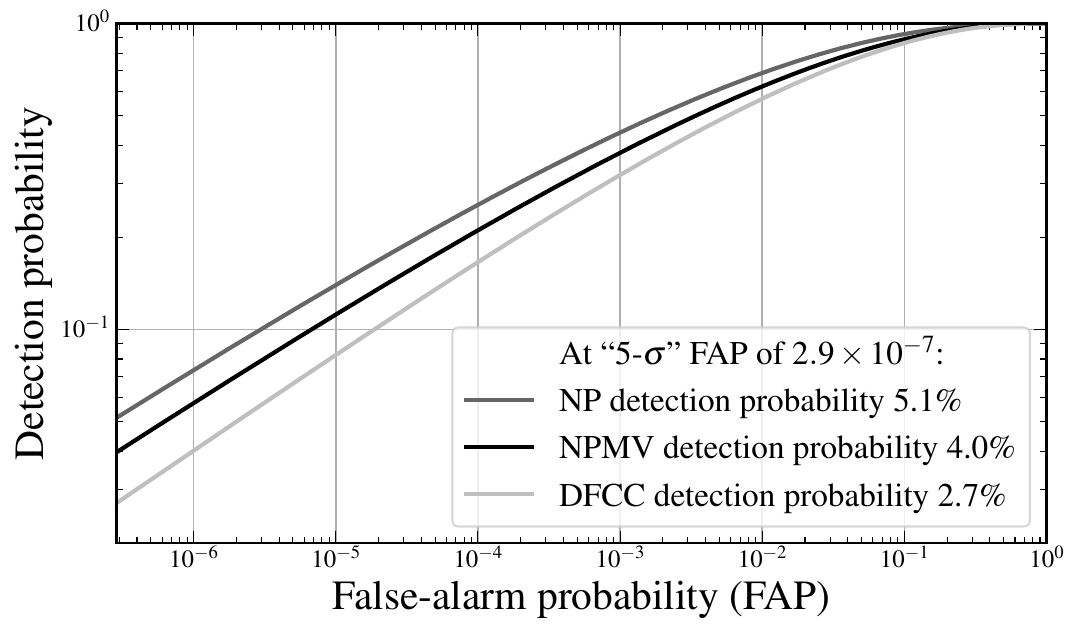}
\caption{ROC curves for the $\DFCC$, $\NP$, and $\NPMV$ detection statistics, 
  for the NANOGrav 15-year PTA fixed $\gamma=13/3$ GWB and pulsar noise
  model~\citep{agazieNANOGrav15Yr2023b}.
  At the discovery threshold FAP of $2.9 \times 10^{-7}$ (leftmost
  plot boundary) the $\NPMV$ statistic increases the detection
  probability by $\sim\!47\%$ compared to the currently standard
  ``optimal" cross-correlation detection statistic $\DFCC$.}
\label{fig:ROC_NG15}
\end{figure}

To compare the different detection statistics in a more realistic
setting, we use the published NANOGrav 15-year ``fixed-$\gamma$''
signal and pulsar noise parameter model values%
\footnote{Exact values of the noise parameters can be found at: 
\url{https://github.com/nanograv/15yr_stochastic_analysis/blob/main/tutorials/data/optstat_ml_gamma4p33.json}.} 
\citep[as used in Fig.~1c of ][]{agazieNANOGrav15Yr2023b}, with $A_{\rm gw}=2.1\times 10^{-15}$ and spectral index 
$\gamma=13/3$ for the GWB.
The amplitude $A_{\rm gw}$ and spectral index $\gamma$ are related to the dimensionless energy-density spectrum 
$\Omega_{\rm gw}(f)$ of the GWB via
\be
\Omega_{\rm gw}(f) = \frac{2\pi^2}{3H^2}A_{\rm gw}^2 f_{\rm yr}^2
\left(\frac{f}{f_{\rm yr}}\right)^{5-\gamma}\,.
\ee
Here, $H \approx 70\,{\rm km}\,{\rm s}^{-1} \,
     {\rm Mpc}^{-1}$ is the present-day Hubble constant, and$f_{\rm
  yr}\equiv 1/{\rm yr}$ 
[see, e.g., Sec.~2.5 in \citep{romanoDetectionMethodsStochastic2017}].  

We construct ROC curves for
the different quadratic detection statistics. 
We form the $\NP$ and $\NPMV$ statistics for realistic data
using~\eqref{e:ump_filter_whitened} and \eqref{e:ump_filter_fromdef}. 
These plots of detection
probability versus FAP are shown in Fig.~\ref{fig:ROC_NG15}.  
At the discovery $5$-$\sigma$ threshold FAP
of $2.9 \times 10^{-7}$, the $\NPMV$ statistic results in a $47\%$
increase in detection probability compared to the standard deflection-optimal
statistic.  The Neyman-Pearson statistic $\NP$ has the best
performance (highest detection probability), but it is not robust because
it makes use of autocorrelations.

\substantialchange{
Note that ``robust'' in this context is not the same as
``conservative''.  By robust we mean that the $p$-value obtained with
a mismodeled noise covariance stays close to the $p$-value that would
be obtained with the correct one, in either direction.  A statistic
is conservative if the error in its $p$-value, when it occurs, tends
to make the mismodeled $p$-value \emph{too large}, so that the
significance of a signal is understated rather than overstated.  In
Sec.~\ref{s:mismodel} we show an example where the $\NP$ statistic is less
robust but more conservative than the $\NPMV$ and $\DFCC$ statistics. We adopt
robustness as our criterion, since our goal is a
statistic whose $p$-values remain meaningful under realistic noise
modeling errors.}


\substantialchange{
\section{Robustness of false alarm probability (\texorpdfstring{$p$}{p}-value) under noise mismodeling}
\label{s:mismodel}

A key constraint in defining the NPMV statistic is that the filter
$\BQ_\NPMV$ vanishes on the (block) diagonal.  This guarantees that
pulsar-timing-residual autocorrelations do not contribute to the
$\NPMV$ statistic. Above, we claim that this makes the $\NPMV$
statistic less ``vulnerable'' to pulsar noise mismodeling errors than
a statistic like $\NP$ that utilizes pulsar autocorrelations.  Here,
we quantify this, comparing the robustness of the $\DFCC$, $\NP$, and
$\NPMV$ statistics under noise mismodeling.

Detection statistics are used for null hypothesis testing.  They
answer the question: how closely do the observed data resemble noise?
More quantitatively: with what probability $p$ might an instance of
the noise give rise to a detection statistic value at least as large as the observed detection statistic value?  In practice, after
selecting a statistic, that statistic is (a) evaluated for the
observed data and then (b) a $p$-value is computed for that observed
statistic value.  Indeed, the \emph{only} use of detection statistics
is to obtain $p$-values.

How does one assess the robustness of a given statistic under noise
mismodeling?  We do this as follows. For many simulated datasets, we
compare the $p$-values obtained with a \emph{correct} noise model to
the $p$-values, obtained for the \emph{same} simulated data, with an
\emph{incorrect} noise model.  Since $p$ values are used to assert
detection claims, these simulated datasets are drawn from an assumed
signal distribution.

The critical quantity is the difference $\epsilon \equiv p_m - p$
between the mismodeled and properly modeled $p$-values.  Useful
diagnostics to quantify the robustness of a detection statistic are
the mean error $\ME$ and the mean-squared error $\MSE$.  Respectively,
these are the mean values of $\epsilon$ and $\epsilon^2$ under the
signal hypothesis.

For a simple PTA toy model, we investigate the probability distribution 
of $\epsilon$, its cumulative distribution, and its mean and second moment.  
This shows clearly that
the $\DFCC$ and $\NPMV$ statistics are significantly more robust than
$\NP$.  A more realistic and systematic treatment of the effects of
pulsar noise mismodeling is an important topic for future
investigation.

\subsection{General framework}
\label{ss:generalframework}

The $p$-value for a detection statistic $D\equiv D(x)$ is defined (see
footnote~\ref{fn:DefinePvalue}) by $p\equiv {\rm Prob}(D>D_{\rm
  obs}\mid H_N)$.  Here, $D_{\rm obs} \equiv D(x_{\rm obs})$ is the
value that the detection statistic takes for the observed data $x_{\rm
  obs}$.  More explicitly, the $p$-value for the observed data is
\be p(x_{\rm obs})\equiv {\rm Prob}\bigl(D(x)>D(x_{\rm obs})\mid x\sim
H_N\bigr)\,,
\label{e:pvals}
\ee
where the probability distribution of $D$ is calculated under the null
hypothesis $H_N$.  For a quadratic detection statistic defined by
filter $\BQ$, the detection statistic is $D(x)\equiv x^T \BQ x$. Note
that in this section we assume that the data $x$ are real-valued.

It is helpful to write the $p$-value \eqref{e:pvals} in terms of a
cumulative distribution function $F$.  This function of three
arguments is defined by \eq{e:def} with the complex integration
variables $z$ replaced by real integration variables $x$, with the
exponent $-z^\dagger \BC^{-1} z$ replaced by $-\frac{1}{2} x^T \BC^{-1}
x$, and with $\det(\pi \BC)$ replaced by $\sqrt{ \det(2 \pi \BC) }$,
so that again $\langle x x^T \rangle = \BC$.  Then, $p$
may be written as
\be p(x_{\rm obs}) = p(x_{\rm obs},\BN,\BQ) \equiv 1- F(x_{\rm obs}^T \BQ x_{\rm obs}, \BN, \BQ)\,.
\label{e:p_CDF}
\ee
This expression for $p$ shows its explicit dependence on (a) the
observed data $x_{\rm obs}$, (b) the noise covariance matrix $\BN$
defining the null hypothesis $H_N$, and (c) the choice of filter
$\BQ$.

As discussed in detail in Sec.~\ref{s:GX2}, the distribution of $D$
under $\BN$ is a generalized $\chi^2$ distribution. This distribution
is completely determined by the eigenvalues of $\BE\equiv
\BN^{1/2}\BQ\BN^{1/2}$.  Thus,
\be
F(x^T \BQ x, \BN, \BQ) = F(x^T \BQ x, \BN^{1/2}\BQ\BN^{1/2}) \,,
\label{e:FtwoVersusFthreeargs}
\ee
where the function on the lhs depends upon three arguments, whereas
that on the rhs depends only upon two.  For generalized $\chi^2$
distributions, there exist several methods to efficiently calculate
the rhs, see, e.g.,~\cite{dasMethodIntegrateClassify2021}.

For a properly modeled analysis, the $p$-value is given by
\eqref{e:p_CDF}.  For a mismodeled analysis, the null-hypothesis
covariance matrix $\BN$ takes the incorrect value $\BN_m$, where
the subscript $m$ denotes ``mismodeled''. The
mismodeled $p$-value is therefore
\be
\begin{aligned}
p_m(x_{\rm obs})
&\equiv p(x_{\rm obs},\BN_m,\BQ_m) \\
&= 1-F\!\left(x_{\rm obs}^T \BQ_m x_{\rm obs},
\BN_m,\BQ_m\right)\\
&= 1-F\!\left(x_{\rm obs}^T \BQ_m x_{\rm obs},
\BN_m^{1/2}\BQ_m \BN_m^{1/2}\right)\,,
\end{aligned}
\label{e:pm_CDF}
\ee
where the first two equalities follows from \eq{e:p_CDF} and the last
from \eq{e:FtwoVersusFthreeargs}.  Here, $\BQ_m$ denotes the
mismodeled filter. It is necessarily mismodeled, because the
(different) detection statistic filters depend upon $\BN$.  So if
$\BN$ is mismodeled then the filter $\BQ$ is also mismodeled.

\subsection{Quantifying the effects of mismodeling}
\label{ss:quantifymismodel}

We quantify the effects of noise mismodeling via
\be
\epsilon(x_{\rm obs}) \equiv
p_m(x_{\rm obs}) - p(x_{\rm obs})\,,
\label{e:Delta_p}
\ee
which is the error in the $p$-values arising from mismodeling.  In
particular, we want to understand which of the different detection
statistics described in this work lead to small $|\epsilon|$ and
which lead to large values.  For this purpose, we cannot make the
comparison for just a single observation $x_{\rm obs}$, because this
leads to conclusions that would depend both upon the data realization 
and upon the statistic.

To make statements purely about the statistic, we study the error in
$p$ for {\it many} realizations of data drawn from an ensemble of
possible universes.  We are interested in the possibility of detecting
a GWB signal, so we characterize the $p$-value errors $\epsilon(x)$
for simulated data drawn from the signal hypothesis $x \sim H_S$,
defined by the properly modeled covariance matrix $\BS$.  To assess
robustness, we define the mean error (ME) and the mean-square error (MSE):
\be
\ba
&\ME \equiv \big\langle \epsilon(x)\big\rangle_{x \sim H_S}\,,
\\
&\MSE \equiv \Big\langle \bigl(\epsilon(x) \bigr)^2\Big\rangle_{x \sim H_S}\,,
\label{e:MSE_def}
\ea
\ee
where the averages are over many realizations $x$ of data drawn from
$H_S$.  The rhs of the above expressions are functions of the
correctly modeled signal covariance matrix $\BS$, as well as $\BN$,
$\BQ$, $\BN_m$, and $\BQ_m$.  However, since $\BN \equiv \Diag \BS$
and the filters $\BQ$ are determined by $\BN$ and $\BS$, the rhs of
\eq{e:MSE_def} is fully determined by $\BS$ and $\BS_m$.

We say that a detection statistic has a small {\it bias} and is {\it
  robust} against noise mismodeling if both $\ME$ and $\MSE$ are
small.  In Sec.~\ref{ss:results_simulations}, we will compare the
$\ME$s and $\MSE$s calculated for the DFCC, NP, and NPMV statistics in
the context of a simple toy model.

\subsection{PTA toy model}
\label{s:toymodel_mismodel_def}

We consider a toy model PTA built from the first $n=10$ pulsars, in
dictionary order from the 67 NANOGrav~15-yr pulsars list: J0023+0923,
J0030+0451, $\dots$, J0613-0200.  We assume that each pulsar is
observed $N_s=100$ times. So a simulated ``observed'' data set for the
entire PTA consists of 1000 samples $x_{ai}$, where $a=1,\dots,n$
labels pulsars, and $i=1,2,\dots, N_s$ labels time samples.  These
samples are drawn from a real zero-mean multivariate Gaussian
distribution.  For a given pulsar $a$, the samples $x_{ai}$ and
$x_{aj}$ are identically distributed, and independent of each other
for $i \ne j$.  (Note, however, that under the signal hypothesis, the
$x_{ai}$ are correlated with $x_{bj}$ for $i=j$, even if $a \ne b$.)

The signal and null hypotheses are defined by $n N_s\times n N_s$
covariance matrices
\be
S_{ai,bj} \equiv S_{ab}\,\delta_{ij}\,,
\qquad
N_{ai,bj} \equiv N_{ab}\,\delta_{ij}\,,
\ee
where the $n\times n$ matrices $\BS$ and $\BN\equiv {\rm diag}(\BS)$
are defined by 
\be
S_{ab} \equiv \ch^2 \, \mu_{ab}+
\sigma^2\, m_a\,\delta_{ab}\,,
\quad
N_{ab} \equiv S_{aa}\,\delta_{ab}\,.
\label{e:mismodel_S_N}
\ee
Here, the HD matrix $\mu_{ab}$ is defined by~\eq{e:HDmat}, $\ch^2$ is
the mean-squared amplitude of the GWB and $\sigma^2$ sets the scale of
the pulsar noise.  For this study, we set $\ch^2=0.64$ and
$\sigma^2=1$.  This yields $p$-values of order $10^{-3}$, similar to
those found in the (updated) NANOGrav
analysis~\cite{vallisneriPosteriorPredictiveChecking2023}.

The $m_a$ are fixed multiplicative factors that we use to simulate
noise-mismodeling errors.  Setting $m_a=1$ for $a=1,2,\dots, 10$
defines the properly modeled covariance matrices $\BS$ and $\BN$.  For
these, all ten pulsars have the same mean-squared noise amplitude
$\sigma^2$.  To mismodel the noise, we set the $m_a \ne 1$. For this
work, we make ten draws from a log-uniform distribution between
$10^{-0.4}\approx 0.4$ and $10^{+0.4}\approx 2.5$ to obtain specific
values of $m_a$ for $a=1,2,...,10$.  (We chose $10^{\mp 0.4}$ for the
limits of the log-uniform distribution to make the effect of noise
mismodeling easy to see.)  In pulsar dictionary order, the values used
in this work are
\be
\begin{aligned}
m_a ={}&
\begin{array}[t]{@{}rrrrr@{}}
  0.703, & 1.480, & 2.454, & 1.238, & 1.383
  , \\
0.683, & 1.408, & 0.576, & 0.650, & 0.490 .
\end{array}
\end{aligned}
\ee
These values for $m_a$, when substituted into \eqref{e:mismodel_S_N},
define the mismodeled covariance matrices $\BS_m$ and $\BN_m$.  From
$\BS$ and $\BN$ (or $\BS_m$ and $\BN_m$), we calculate
properly modeled (or mismodeled) filters $\BQ$ (or $\BQ_m$) that are
then used to define the corresponding quadratic detection statistics
for DFCC, NP, and NPMV.
\begin{figure}[hbtp!] 
\centering
\includegraphics[width=0.98\linewidth]{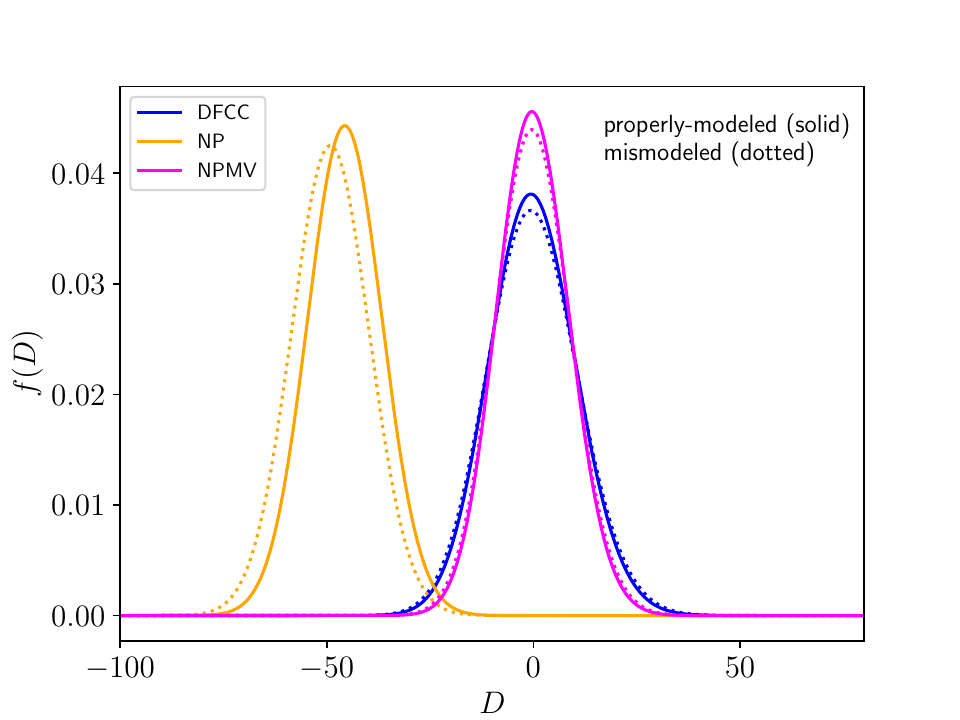}
\caption{\substantialchange{The probability distributions $f(D)$ for the
  $\DFCC$, $\NP$, and $\NPMV$ detection statistics $D$ under the null
  hypothesis are generalized $\chi^2$.
  Solid curves: correctly modeled; dotted curves: mismodeled.
Normalization $\int f(D) \, {\rm d} D = 1$.
}}
\label{f:mismodel_GX2} 
\end{figure}

As discussed above, the distributions of the three different detection
statistics are generalized $\chi^2$ distributions, determined by the
eigenvalues of the matrix $\BE\equiv \BN^{1/2}\BQ\BN^{1/2}$.  These
eigenvalues may be computed exactly; the resulting distributions are
shown in Fig.~\ref{f:mismodel_GX2} for the three different detection
statistics.  For each detection statistic $D=x^T \BQ x$, we show the
distribution of $D$ for the correct null hypothesis $\BN$ and
corresponding correct filter $\BQ$, and for the mismodeled null
hypothesis $\BN_m$ and corresponding (mismodeled) filter $\BQ_m$.

The distributions for the $\DFCC$ and $\NPMV$ statistics under either
$\BN$ or $\BN_m$ have zero mean: the expected value of $D(x)$ vanishes
under the null hypothesis. In contrast, the mean of the $\NP$
statistic is negative for both the properly and incorrectly modeled
cases.  For all three statistics, the variance under the incorrectly
modeled hypothesis is slightly larger than under the correct
hypothesis.

\subsection{Results of numerical simulations}
\label{ss:results_simulations}

To study the errors $\epsilon(x_{ai}) \equiv p_m(x_{ai}) - p(x_{ai})$
in $p$-values arising from mismodeling in the toy model, we simulate
data $x_{ai}$ under the signal hypothesis. A single data set is
simulated as follows.

To create samples of zero-mean Gaussian random variables with the
desired correlations $\BS$ across pulsars $a$, $b$, we follow the
standard eigenvector-eigenvalue decomposition method, as described,
e.g., in \cite{Renzini2023}.  Explicitly, we decompose $\BS$ as $\BS =
\tilde\BE\boldsymbol{\Lambda}\tilde\BE^T$, where $\tilde\BE$ is the
matrix having the normalized eigenvectors of $\BS$ as columns, and
$\boldsymbol{\Lambda}$ is the diagonal matrix having the eigenvalues
$\lambda_b$ of $\BS$ along the diagonal.  Then for each $i=1,2,\ldots,
N_s$, we define $x_{ai} \equiv\sum_b \tilde
E_{ab}\sqrt{\lambda_b}\,z_{bi}$, where $z_{bi}$ are $nN_s$ independent
samples drawn from a zero-mean unit-variance Gaussian distribution.
\begin{table}
\substantialchange{
  \begin{ruledtabular}
\begin{tabular}{
  r
  r@{\hspace{0.15em}$\times\!$}l
  r@{\hspace{0.15em}$\times\!$}l
  r@{\hspace{0.15em}$\times\!$}l
  r@{.}l
  r
}
Statistic
  & \multicolumn{2}{c}{$\langle p\rangle$}
  & \multicolumn{2}{c}{$\ME = \langle\epsilon\rangle$}
  & \multicolumn{2}{c}{$\MSE = \langle\epsilon^{2}\rangle$}
  & \multicolumn{2}{c}{$\frac{\ME}{\langle p\rangle}$}
& $\frac{\MSE}{\langle p\rangle^2}$ \\
\noalign{\vskip 1pt}
\hline
\noalign{\vskip 1pt}
$\DFCC$
  & $6.0$ & $10^{-4}$
  & $1.4$ & $10^{-4}$
  & $1.6$ & $10^{-5}$
  & $0$ & $\!\!23$
  & $44$ \\
$\NP$
  & $4.0$ & $10^{-4}$
  & $16.$ & $10^{-4}$
  & $14.$ & $10^{-5}$
  & $4$ & $\!\!0$
  & $875$ \\
$\NPMV$
  & $5.3$ & $10^{-4}$
  & $1.4$ & $10^{-4}$
  & $1.8$ & $10^{-5}$
  & $0$ & $\!\!26$
  & $64$ \\
\end{tabular}
\end{ruledtabular}
\caption{\label{t:meanpeps}\substantialchange{The mean values of $p$, $\epsilon \equiv
  p_m-p$, and $\epsilon^2$ for $10^8$ simulations of the toy
  model. The $\DFCC$ statistic is the most robust, followed closely
  behind by $\NPMV$. The $\NP$ statistic is the least robust, having
  much larger $p$-value mismodeling errors than $\DFCC$ and $\NPMV$.
  For example, $\NP$'s MSE is $\sim 14$~-~$20$ times larger than for the
  other statistics (final column).}}
}
\end{table}

To evaluate the robustness of the three different detection statistics
for the toy model, we generate $10^8$ realizations of the PTA data
$x_{ai}$ under the signal hypothesis $H_S$, and compute $p(x_{ai})$
and $p_m(x_{ai})$ for each of these realizations, for the three
different statistics.  Thus, the complete output of the simulation is
a file containing $10^8$ rows. Each row contains three correctly
modeled $p$-values, and three incorrectly modeled $p$-values.  The
mean of (the correctly modeled) $p$ for the three different detection
statistics are given in Table~\ref{t:meanpeps}.  Note that we
deliberately selected the signal-model GWB intensity $\ch^2 = 0.64$ so
that the mean $\DFCC$ $p$-value $\langle p_\DFCC \rangle \approx 6.0 \times
10^{-4}$ corresponds to the $3.2\sigma$ significance found in the
(updated) NANOGrav
analysis~\cite{vallisneriPosteriorPredictiveChecking2023}.

\begin{figure*}[!t]
\centering
\includegraphics[width=0.98\textwidth]{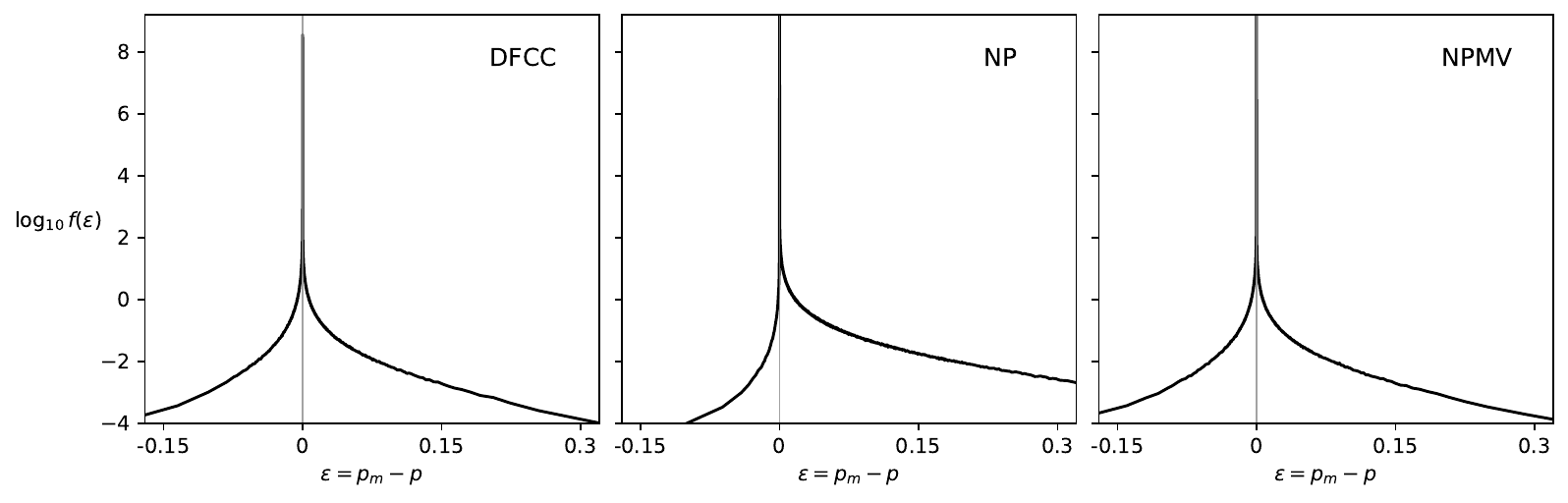}
\caption{\substantialchange{The probability density $f(\epsilon)$ of the errors $\epsilon
  \equiv p_m - p$ arising from noise mismodeling in our toy model,
  normalized so that the area under the curve is unity: $\int_{-1}^{1}
  f(\epsilon) {\rm d} \epsilon =1$.  The spikes are like Dirac
  $\delta$-functions: the region $- \langle p \rangle/2 < \epsilon <
  \langle p \rangle/2$ contains $\sim 90\%$ of the probability for
  $\DFCC$ and $\NPMV$, and $\sim 80\%$ of the probability for
  $\NP$. See Fig.~\ref{f:CDF} for the CDFs.
  \label{f:edist} }}
\end{figure*}

\begin{figure*}[!t]
\centering
\includegraphics[width=0.98\textwidth]{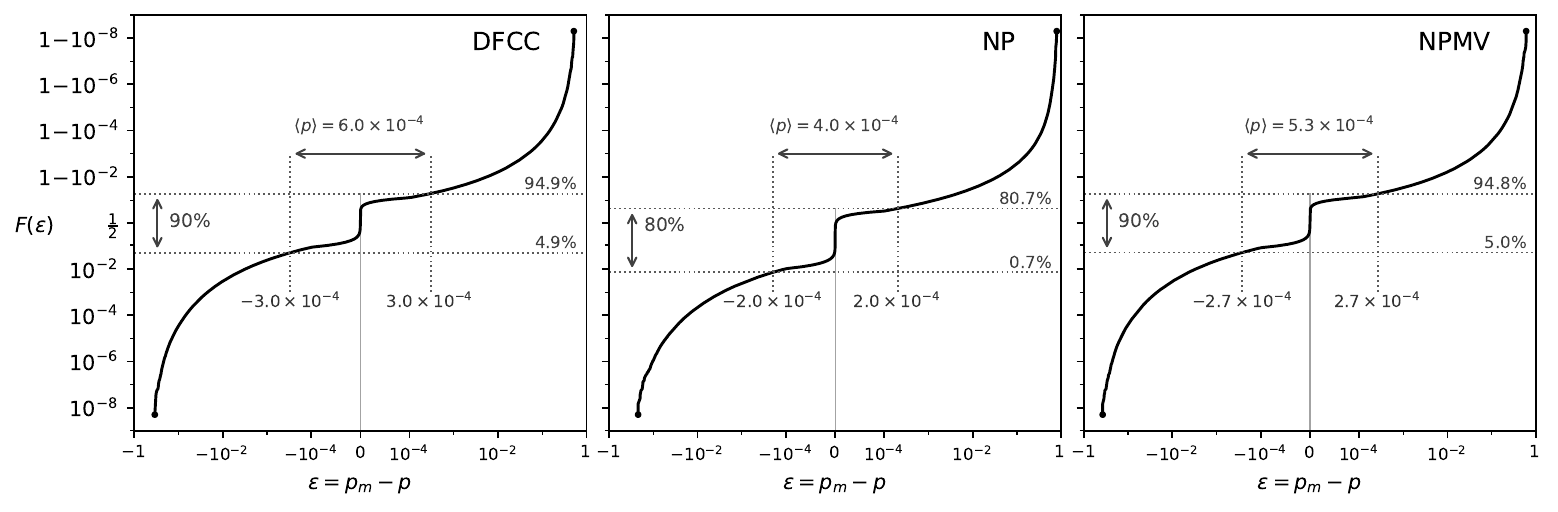}
\caption{\substantialchange{The cumulative distribution functions (CDF) of the errors in
  the three statistics, defined by $F(\epsilon) = \int_{-1}^\epsilon
  f(x) {\rm d}x$. The dots denote the largest and smallest $\epsilon$
  values among the $10^8$ samples.  For the three statistics,
  $80$--$90\%$ of signal-containing universes produce insignificant
  errors $\epsilon$ in $p$.  The remaining $10$--$20\%$ are
  problematic: the mismodeled $p$-value can be orders of magnitude too
  large.  In the case of $\DFCC$ and $\NPMV$, about 5\% of the
  $p_m$-values are significantly too small ($\epsilon<0$) and about
  5\% are significantly too large ($\epsilon>0$).  In the case of
  $\NP$, less than 1\% of $p_m$ are significantly too small, and more
  than $19\%$ are significantly too large. This is why $\NP$ is less
  robust than the other statistics.}}
\label{f:CDF}
\end{figure*} 

\begin{figure*}[!t]
\centering
\includegraphics[width=0.98\textwidth]{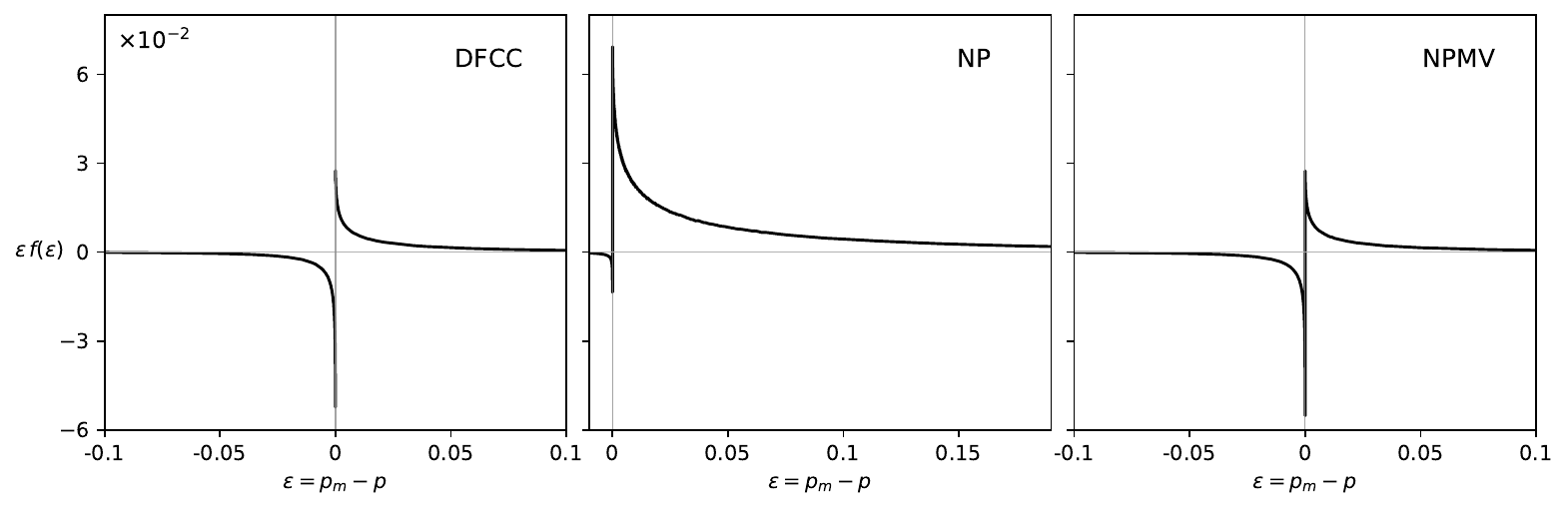}
\caption{\substantialchange{Contributions to the mean error (ME) or bias between the
  $p$-values for properly and improperly modeled noise: the area under
  these curves is the $\ME$.  The $\delta$-function-like spikes make
  no significant contribution: what dominates is the small fraction of
  $p$-value estimates which are incorrect by orders of magnitude. The
  $\DFCC$ and $\NPMV$ errors have both signs, which reduces the
  average bias. In contrast, the mismodeled $\NP$ $p$-value is almost
  always too large, so the $\NP$ bias is much larger.}}
\label{f:FirstMoment}
\end{figure*}

\begin{figure*}[!t]
\centering
\includegraphics[width=0.98\textwidth]{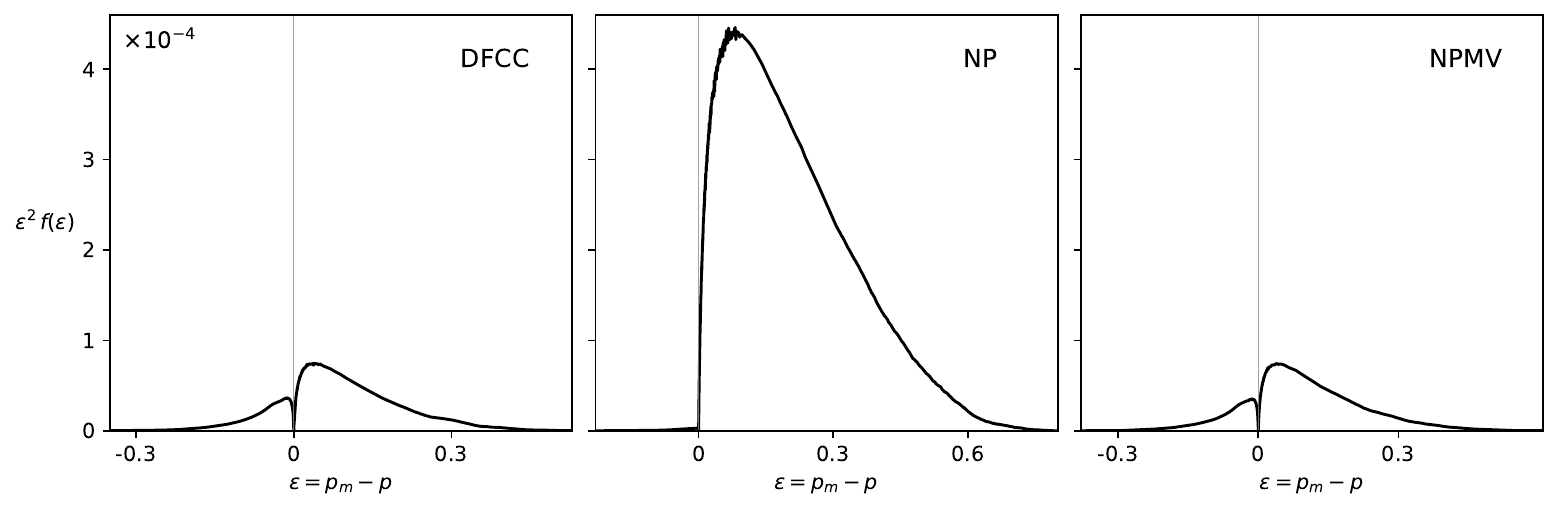}
\caption{\substantialchange{Contributions to the mean squared error (MSE) between the
  $p$-values for properly and improperly modeled noise.  The area
  under this curve is the $\MSE$.  Again, the dominant contribution is
  from the small fraction of $p$-value estimates which are wildly
  incorrect.  Under improper noise modeling, the $\NP$ statistic is
  significantly more likely to overestimate the false alarm
  probability and it does so about twice as often as the other
  statistics.}}
\label{f:SecondMoment}
\end{figure*}

From the simulation output, we compute the mismodeling errors
$\epsilon \equiv p_m - p$ for each of the three detection statistics.
Shown in Fig.~\ref{f:edist} are the probability densities
$f(\epsilon)$ of these errors $\epsilon$. They are sharply peaked at
zero, because most signal realizations give rise to $p$-value errors
that are comparable to the mean $p$-value, of order $10^{-3}$.  As can
be seen from the cumulative distributions, shown in Fig.~\ref{f:CDF},
$80$--$90\%$ of the mismodeled $p$-values have negligible errors:
$\epsilon \in [-1, 1] \langle p \rangle/2$.  But there
are some signal realizations where the properly modeled $p$-value is
close to the correct value $O(10^{-3})$, and the incorrectly modeled
value is of order unity, three orders of magnitude larger!  These
rare cases have a dramatic effect.

From the simulation output, we also compute the mean of the
mismodeling errors, $\ME \equiv \langle \epsilon \rangle$, and the
mean of the squares, $\MSE \equiv \langle \epsilon^2 \rangle$.  These
are given in Table~\ref{t:meanpeps}. The $\delta$-function-like spikes
visible in Fig.~\ref{f:edist}, which contain $80-90\%$ of the
probability, make no significant contribution to the $\ME$ or $\MSE$.
Rather, these moments are dominated by the $10\%$ of universes (for
$\DFCC,\NPMV$) which have $| \epsilon | \gtrsim \langle p \rangle/2$.
For $\NP$ this fraction is $20\%$, and the impact is significantly
greater.

The behavior is very different for the ``robust'' statistics ($\DFCC$,
$\NPMV$) and for the ``non-robust'' Neyman-Pearson $\NP$ statistic,
where the mismodeling leads to much larger errors.  To understand the
source of these $p$-value errors, it is very helpful to examine the
contributions to the ME, $\langle \epsilon \rangle = \int_{-1}^1
\epsilon f(\epsilon)\,{\rm d}\epsilon$, arising from different values
of $\epsilon$.  Examination of the integrand, plotted in
Fig.~\ref{f:FirstMoment}, shows that these moments are dominated by
the rare values of $p_m$ that are off by orders of magnitude.  In the
case of $\DFCC$ and $\NPMV$ (see Table~\ref{t:meanpeps}) the $\ME$,
which can be interpreted as an average bias, is about a factor of four
\emph{smaller} than the mean $p$-value, whereas for $\NP$ it is about
a factor of four \emph{larger} than the mean $p$-value. In this sense
the robust statistics give unbiased $p$-values, whereas the non-robust
$\NP$ statistic gives biased $p$-values.
Since $\ME > 0$, the $\NP$ bias is conservative (the mismodeled
$p$-value is systematically too large), but a $p$-value that is
systematically wrong in a known direction is still wrong: it is
robustness, not conservativeness, that we want.

Figure~\ref{f:SecondMoment} shows the contributions to the MSE,
$\langle \epsilon^2 \rangle = \int_{-1}^1 \epsilon^2f(\epsilon)\,{\rm
  d} \epsilon$, arising from different values of $\epsilon$. The
situation is the same as described above: for most realizations, the
mismodeling errors $\epsilon$ are small compared to the expected value
of $p$ and thus unimportant. The important contributions come from the
tail of large errors.  The $\DFCC$ and $\NPMV$ statistics are far more
robust than $\NP$, whose $\MSE$ are $14$--$20$ times larger, measured
with respect to the square of the mean $p$-value, $\langle
p\rangle^2$.  See the final column of Table~\ref{t:meanpeps}.

In closing this section, we emphasize that these illustrative results
are for \emph{one} definition of robustness, for \emph{one} toy model,
and for \emph{one} choice of GWB signal/noise/mismodel parameter
values.  Experiments with other values usually give similar $\DFCC$,
$\NP$, and $\NPMV$ detection statistic robustness rankings, but this
is {\em not universal}.  Namely, there exist cases---usually when the
overall amplitude of the noise mismodeling is relatively small and the
GWB is relatively large---when the $\NP$ statistic is more robust than
the $\DFCC$ and $\NPMV$ statistics.  So, more realistic studies of the
effects of pulsar noise mismodeling would be useful.

}

\section{Conclusion}
\label{s:concl}

This paper introduces a new quadratic detection statistic for pulsar
timing arrays (PTA), which we call the
``Neyman-Pearson-Minimum-Variance'' statistic and denote with $\NPMV$.
We argue that the $\NPMV$ statistic should replace the traditional
``optimal" cross-correlation detection statistic, which we denote by
$\DFCC$, and which is currently used by all of the PTA collaborations
to assess their false alarm probabilities (FAPs) and $p$-values.

Like $\DFCC$, the $\NPMV$ statistic uses only pulsar
cross-correlations, making it robust against uncertainties in pulsar
timing residual noise modeling.  But the $\NPMV$ statistic has a
significant advantage over $\DFCC$: at a given FAP, it is more
sensitive.  In simulations that model the current generation of PTAs,
$\NPMV$ has a 47\% higher detection probability than $\DFCC$, when
operating at the $5\sigma$-equivalent FAP of $2.9 \times 10^{-7}$.

The related $\NPCC$ statistic is guaranteed to perform even better
than $\NPMV$.  However, it can only be obtained numerically, and our
investigations indicate that in practice it only provides modest
improvements over $\NPMV$.

The $\NPMV$ filter is implemented in the
\texttt{enterprise\_extensions}
package~\citep{ellisENTERPRISEEnhancedNumerical2020,taylorEnterprise_extensions2021}. This
means that existing analyses based on the $\DFCC$ statistic can be
easily redone using the $\NPMV$ detection statistic.

\begin{acknowledgments}
  We thank Wang-Wei Yu for helpful discussions\substantialchange{, and the anonymous
  referee for motivating us to add Secs.~\ref{ss:numeval} and
  \ref{s:mismodel} and the Appendix}. J.D.R.~acknowledges financial support from the
  National Science Foundation (NSF) Physics Frontier Center Award
  PFC-2020265 and start-up funds from the University of Texas Rio
  Grande Valley.  J.D.R.~performed part of this work at the Aspen
  Center for Physics, which is supported by NSF grant PHY-2210452.
\end{acknowledgments}

\section*{Data availability statement}
No open or closed data were used in this work. We make use of pulsar
positions and model parameters as described in publications in the
literature~\citep{agazieNANOGrav15Yr2023b}. \substantialchange{The ROC curves shown in
Figs.~\ref{fig:ROC_TOY} and~\ref{fig:ROC_NG15}} can be regenerated
using code available online at
\url{https://github.com/vhaasteren/robust-ds-figures}.

\appendix

\begin{widetext}

\substantialchange{
\section{Analytical calculation of the mean error $\ME$ and 
mean-squared error $\MSE$ for Gaussian-distributed detection
statistics}
\label{s:analytic_gauss}

In Sec.~\ref{ss:quantifymismodel} we introduce two measures of the
$p$-value errors that arise from mismodeling. The $\ME$ is the
average, over many signal realizations, of the difference between the
correctly and incorrectly modeled $p$ value, and the $\MSE$ is the
average of its square.  The evaluation of those quantities for a PTA
toy model, was carried out numerically in
Sec.~\ref{ss:results_simulations}. In this Appendix, we investigate a
set of three approximations which yield explicit analytic closed forms for
the $\ME$ and $\MSE$.

The analytic treatment below makes three approximations.
  First, we assume that the probability distributions of the different
  detection statistics under the null hypothesis are Gaussian.  These
  probability distributions are actually generalized $\chi^2$
  distributions; an example is shown in Fig.~\ref{f:mismodel_GX2}.
  This approximation replaces them with Gaussian distributions whose
  mean and variance match those of the generalized $\chi^2$.  Second,
  we assume that the difference between the correctly and incorrectly
  modeled parameters is small enough that the change in the $p$-value
  is accurate at linear order in that difference.  Third, under the
  signal hypothesis, we assume that the squared-data matrix $\BX$
  (which has a Wishart distribution) is accurately represented by a
  joint Gaussian distribution having the correct mean and covariance.

The validity of the first approximation depends upon both
  the order of magnitude of the $p$-value and the spectrum
  of the eigenvalues which
  define the generalized $\chi^2$ distribution.  A generalized
  $\chi^2$ distribution with sufficiently large effective rank is
  often well-approximated near its peak by a Gaussian distribution,
  but the approximation can be poor when a few eigenvalues dominate
  and generally deteriorates far out in the tails.  For the models
  considered here, the analytical Gaussian results give approximations
  to within tens of percent for relatively large $p$-values (e.g.,
  $p\gtrsim 10^{-2}$).  For smaller $p$-values (e.g., $p\lesssim
  10^{-3}$), like those considered in Sec.~\ref{s:mismodel}, they
  differ from the numerical generalized-$\chi^2$ results by factors of
  order unity.

\subsection{Gaussian-distributed detection statistics}
\label{s:gauss-det-stats}

In what follows, we assume that the detection statistic $D(x)$ is
accurately described by a Gaussian distribution under the null
hypothesis $H_N$.  This means that its $p$-value for observed data $x$
may be written as
\be
p(x)= 
\frac{1}{2}{\rm erfc}\left(\frac{ z(x)}{\sqrt{2}}\right)\,,
\qquad
z(x) \equiv\frac{D(x) - \mu}{\sigma}\,,
\label{e:p_gauss_approx}
\ee
where the complementary error function is
\be
{\rm erfc}(u) \equiv\frac{2}{\sqrt{\pi}}\int_u^\infty \dd x\, e^{-x^2} \, .
\label{e:erfc}
\ee
As explained above, in this approximation we set the Gaussian's mean
and variance to match those of the detection statistic:
\be
\mu \equiv \langle D\rangle_{H_N}\,,
\quad
\sigma^2 \equiv \langle D^2\rangle_{H_N}-\langle D\rangle_{H_N}^2\,.
\label{e:mean-variance}
\ee
With these assumptions, it is straightforward to compute how $p$
changes under mismodeling.

For quantifying robustness, we are interested in changes in the
$p$-values due to noise mismodeling (see Secs.~\ref{s:mismodel}
and~\ref{ss:quantifymismodel} for details):
\be
\epsilon(x) \equiv \Delta p (x) \equiv  p_m(x)-p(x)\,.
\ee
The mismodeling affects the $p$-value because it modifies the mean
$\mu$ and variance $\sigma^2$ of the detection statistic under
the null distribution, and because it
changes the value of $D(x)$.  Via \eqref{e:p_gauss_approx}, this
modifies $z$ and hence $p$.  Starting from \eqref{e:p_gauss_approx}
for $p(x)$, we can derive a simple expression for the change $\Delta
p(x)$ in the $p$-value due to changes in $\mu$, $\sigma$, and $D$ for
a fixed data $x$.  Differentiating \eq{e:p_gauss_approx} and using
\be \frac{\dd}{\dd u}\,{\rm erfc}(u) = -\frac{2}{\sqrt{\pi}}
e^{-u^2}\,,
\label{e:erfc'}
\ee
we obtain
\be
\Delta p(x) = \frac{1}{\sqrt{2\pi}\sigma}
e^{-\frac{1}{2}z^2(x)}
\Bigl(\Delta \mu + z(x)\,\Delta\sigma - \Delta D(x)\Bigr)\,,
\ee
where the terms in parenthesis come from applying the chain rule to
\eq{e:p_gauss_approx}, and we drop terms of second or higher order.
Thus, this expression gives $\epsilon$ to first order in the
parameter mismatches.

To compare the performance of different detection statistics when
averaged over many realizations of data as in \eq{e:MSE_def}, we
calculate the mean and mean-squared values of these mismodeling
errors:
\be
{\rm ME}\equiv \langle \epsilon(x)\rangle_{x\sim H_S}\,,
\quad
{\rm MSE}\equiv \langle \epsilon^2(x)\rangle_{x\sim H_S}\,,
\ee
where the ensemble average is over data $x$ drawn from the signal hypothesis
$H_S$.

To obtain explicit expressions for the ME and MSE due to noise
mismodeling, we restrict attention now to quadratic detection
statistics, which are completely described by a quadratic filter $\BQ$
and covariance matrices $\BN$ and $\BS$, as discussed in
Sec.~\ref{ss:generalframework}.  We further assume that the vectors
$\boldsymbol{x}_i\equiv(x_{1i},\ldots,x_{ni})^T$, for
$i=1,\ldots,N_s$, are independent and identically distributed,
zero-mean multivariate Gaussians with covariance $\BS$, as in our
toy model example in Sec.~\ref{s:toymodel_mismodel_def}.  In terms of
$x$, the quadratic detection statistic then takes the form
\be
D(x)
\equiv \sum_{a,b=1}^n\sum_{i=1}^{N_s}x_{ai} Q_{ab} x_{bi} ={\rm Tr}(\BQ\BX)\,,
\label{e:D=Tr(QX)}
\ee
where the data $\BX \equiv X_{ab}$ is a symmetric $n \times n$ 
dimensional matrix
\be
X_{ab}\equiv \sum_{i=1}^{N_s} x_{ai}x_{bi}\,.
\label{e:X_ab}
\ee
Under the signal hypothesis $H_S$, the data matrix $\BX$ has mean and covariance
\be
\begin{aligned}
&\BM \equiv M_{ab}\equiv \langle X_{ab}\rangle_{H_S} = N_s S_{ab}\,,
\\
&\BK \equiv K_{ab,cd}\equiv \langle X_{ab}X_{cd}\rangle_{H_S} -
\langle X_{ab}\rangle_{H_S} 
\langle X_{cd}\rangle_{H_S} 
= N_s (S_{ac} S_{bd} + S_{ad} S_{bc})\,.
\label{e:MK}
\end{aligned}
\ee
To see explicitly why these moments are exact, Isserlis'
theorem~\cite{isserlisFormulaProductmomentCoefficient1918} and
independence of samples with different $i$ give
\be
\left\langle
x_{ai}x_{bi}x_{cj}x_{dj}
\right\rangle_{H_S}
=S_{ab}S_{cd}
+\delta_{ij}\left(S_{ac}S_{bd}+S_{ad}S_{bc}\right)\,.
\label{e:Isserlis_X}
\ee
Summing over $i$ and $j$ therefore yields
\be
\left\langle X_{ab}X_{cd}\right\rangle_{H_S}
=N_s^2S_{ab}S_{cd}
+N_s\left(S_{ac}S_{bd}+S_{ad}S_{bc}\right)\,,
\label{e:XX_exact}
\ee
and subtracting
$\langle X_{ab}\rangle\langle X_{cd}\rangle
=N_s^2S_{ab}S_{cd}$ gives \eqref{e:MK}.

For Gaussian sample vectors $\boldsymbol{x}_i$, the full matrix
$\BX=\sum_i\boldsymbol{x}_i\boldsymbol{x}_i^T$ has the Wishart
distribution $W_n(\BS,N_s)$.  A single component $X_{ab}$ is a
generalized $\chi^2$ random variable; in particular,
$X_{aa}/S_{aa}\sim\chi^2_{N_s}$.  The matrix $\BX$ is not itself
Gaussian-distributed for finite $N_s$.

The mean and covariance in \eqref{e:MK} are therefore exact even
though $\BX$ is not Gaussian.  In the $\ME$ and $\MSE$ integrals below, we
approximate the components of the Wishart
matrix $\BX$ by a joint Gaussian distribution with these exact
moments.  Similarly, under the null hypothesis, where the sample
vectors have covariance $\BN$, \eqref{e:D=Tr(QX)} gives the exact mean
and variance
\be
\begin{aligned}
&\mu\equiv \langle D\rangle_{H_N} = N_s{\rm Tr}(\BQ\BN)\,,\\
&\sigma^2 \equiv \langle D^2\rangle_{H_N}-\langle D\rangle_{H_N}^2
= 2N_s{\rm Tr}(\BQ\BN\BQ\BN)\,.
\label{e:mu_sigma}
\end{aligned}
\ee
Equations \eqref{e:D=Tr(QX)}, \eqref{e:MK}, and 
\eqref{e:mu_sigma} 
are needed for the analytic ME and MSE calculations
that follow.

\subsection{Calculation of the mean error}

We begin by calculating the mean error due to noise
mismodeling, which is the average of 
$\epsilon(\BX)$ with respect to the 
Gaussian probability distribution for $\BX$ under 
$H_S$ (see \eqref{e:MK}):
\be
{\rm ME}
\equiv 
\langle \epsilon(\BX)\rangle_{\BX\sim H_S}
\equiv
\int \dd \BX \>
\frac{1}{\sqrt{{\rm det}(2 \pi \BK)}}
{\rm exp}\left[-\frac{1}{2}(X_\alpha-M_\alpha)
K^{-1}_{\alpha\beta}(X_\beta-M_\beta)\right]\,
\epsilon(\BX)\,,
\label{e:ME_int}
\ee
where
\be
\epsilon (\BX) = \frac{1}{\sqrt{2\pi}\sigma}
e^{-\frac{1}{2}z^2(\BX)}
\Bigl(\Delta \mu + z(\BX)\,\Delta\sigma - \Delta D(\BX)\Bigr)\,.
\label{e:eps_def}
\ee
In \eqref{e:ME_int}, the repeated Greek indices $\alpha$ and $\beta$
are summed over, and 
$X_\alpha$, $M_\alpha$, and $K_{\alpha\beta}$
are {\it symmetry-reduced versions} of the symmetric matrices 
$X_{ab}$, $M_{ab}$, and $K_{ab,cd}$.
The indices $\alpha$, $\beta$ run over values $1,2,\ldots, m$, 
where $m=n(n+1)/2$ is the number of independent components of 
a symmetric $n\times n$ matrix.
When we need to be explicit, we will write $\alpha\equiv (i,j)$, for $1\le i\le j\le n$.

We assume that $\BS$ is positive definite, so the symmetry-reduced
covariance $\BK$ is nonsingular, and define the integration measure by
\be
\dd\BX\equiv\prod_{\alpha=1}^{m}\dd X_\alpha\,.
\label{e:dX_measure}
\ee
The quantities $X_\alpha$, $M_\alpha$, $Q_\alpha$, and
$K^{-1}_{\alpha\beta}$
appearing in \eqref{e:ME_int} 
can be thought of as the components of $m$-dimensional vectors $\BX$
and $\BM$, a corresponding reduced filter vector, and the inverse of
the $m\times m$ matrix $\BK$ defined via:
\be
X_\alpha \equiv E_{ab}^\alpha X_{ab}\,,
\qquad
M_\alpha \equiv E_{ab}^\alpha M_{ab}\,,
\qquad
K_{\alpha\beta} \equiv E_{ab}^\alpha K_{ab,cd} E_{cd}^\beta\,,
\ee
and
\be
\begin{aligned}
Q_\alpha&\equiv E^\alpha_{ab}Q_{ab}\,,
&
Q_{m,\alpha}&\equiv E^\alpha_{ab}Q_{m,ab}\,,
&
\Delta Q_\alpha&\equiv Q_{m,\alpha}-Q_\alpha\,.
\end{aligned}
\label{e:Qalpha_defs}
\ee
where (again) repeated indices are summed over, and 
$E^\alpha_{ab}$ for $\alpha=1,2\ldots, m$ 
comprise an orthonormal and complete set
of basis ``vectors" for the space of symmetric $n\times n$ matrices.
The orthonormality and completeness conditions are:
\be
E_{ab}^\alpha E_{ab}^\beta = \delta_{\alpha\beta}\,,
\qquad
E_{ab}^\alpha E_{cd}^\alpha 
= \frac{1}{2}\left(\delta_{ac}\delta_{bd} + \delta_{ad}\delta_{bc}\right)
\equiv I_{ab,cd}\,.
\label{e:orthonormal-complete}
\ee
Note that $I_{ab,cd}$ are the components of the identity matrix restricted
to acting on symmetric $n\times n$ matrices.
Explicit expressions for the $E^\alpha_{ab}$ are
\be
E^\alpha_{ab}\equiv E^{(ij)}_{ab} 
= \begin{cases}
\delta_{ai}\delta_{bi} & i=j\,,\\
\frac{1}{\sqrt{2}}\left(\delta_{ai}\delta_{bj} + \delta_{aj}\delta_{bi}\right) & i<j\,,
\end{cases}
\ee
where the $1/\sqrt{2}$ normalization factor ensures 
that \eqref{e:orthonormal-complete} holds.

For positive-definite $\BS$, the inverse covariance restricted to the
space of symmetric matrices is
\be
(K^{-1})_{ab,cd}
=\frac{1}{4N_s}\left[
(S^{-1})_{ac}(S^{-1})_{bd}
+(S^{-1})_{ad}(S^{-1})_{bc}\right]\,,
\label{e:Kinv_full}
\ee
which satisfies
\be
K_{ab,cd}(K^{-1})_{cd,ef}
=I_{ab,ef}
=\frac{1}{2}\left(
\delta_{ae}\delta_{bf}+\delta_{af}\delta_{be}\right)\,.
\label{e:Kinv_full_identity}
\ee
Its symmetry-reduced form is
\be
K^{-1}_{\alpha\beta}
\equiv E^\alpha_{ab}(K^{-1})_{ab,cd}E^\beta_{cd}\,,
\qquad
K_{\alpha\beta}K^{-1}_{\beta\gamma}
=\delta_{\alpha\gamma}\,.
\label{e:Kinv_reduced}
\ee
As a consistency check, we can derive a formula for the determinant of
the operator $\BK$ in terms of the determinant of $\BS$. For this,
diagonalize $\BS$, and denote its eigenvalues by $\lambda_i>0$.  On the
space of symmetric matrices $\BY$, $\BK$ acts as $\BK (\BY )=2N_s \BS
\BY \BS$.  Consequently, if we work in coordinates in which $\BS$ is
diagonal, then the diagonal basis matrices $E^{(ii)}$ for $\BY$ have eigenvalues
$2N_s \lambda_i^2$, and the off-diagonal basis matrices $E^{(ij)}$,
with $i<j$, have eigenvalues $2N_s \lambda_i \lambda_j$.  Multiplying
all $m=n(n+1)/2$ eigenvalues, each $\lambda_i$ occurs with total power
$2+(n-1)=n+1$, giving
\be
\det\BK=(2N_s)^m(\det\BS)^{n+1}\,.
\label{e:detK}
\ee
Since $\BS$ is assumed positive definite, our explicit construction of
its eigenvalues proves that $\BK$ is also positive definite.

The orthonormality conditions \eq{e:orthonormal-complete} also ensure
that inner products are preserved, so that e.g., $Q_{ab} X_{ab} =
Q_\alpha X_\alpha \equiv \BQ^T \BX$ and
\be
\qquad
(X_{ab}-M_{ab}) K^{-1}_{ab,cd} (X_{cd}-M_{cd}) =
(X_{\alpha}-M_{\alpha}) K^{-1}_{\alpha\beta} (X_\beta-M_\beta)
\equiv 
(\BX-\BM)^T \BK^{-1}(\BX-\BM)\,.
\ee
In terms of the above symmetry-reduced vectors and matrices, we can 
write \eqref{e:ME_int} in a more compact (index-free) form as:
\be
{\rm ME}
= \int \dd \BX \>
\frac{1}{\sqrt{{\rm det}(2 \pi \BK)}}
{\rm exp}\left[-\frac{1}{2}(\BX-\BM)^T\BK^{-1}(\BX-\BM)\right]\,\epsilon(\BX)\,,
\label{e:ME_int2}
\ee
where $(\BX-\BM)^T$ denotes the transpose of the vector $\BX-\BM$.

The quantities $\Delta\mu$, $\Delta\sigma$, and $\Delta D(\BX)$, 
which enter $\epsilon(\BX)$, are obtained by evaluating $\mu$, $\sigma$,
and $D(\BX)$ first for the mismodeled covariance matrices and filter
and then subtracting the corresponding quantities using the properly modeled
covariance matrices and filter.
Explicitly, with $\sigma>0$ and $\sigma_m>0$, our sign convention is
\be
\Delta\mu\equiv\mu_m-\mu\,,
\qquad
\Delta\sigma\equiv\sigma_m-\sigma\,,
\qquad
\Delta Q_\alpha\equiv Q_{m,\alpha}-Q_\alpha\,,
\qquad
\Delta D(\BX)\equiv D_m(\BX)-D(\BX)\,.
\label{e:Delta_convention}
\ee
The data $\BX$ do not change in this noise-mismodeling 
calculation since the data are always drawn from the properly modeled 
signal covariance $\BS$.
The data enter $\epsilon(\BX)$ via $z(\BX)$ and $\Delta D(\BX)$, which 
have the explicit form:
\be
z(\BX) 
= \frac{D(\BX)-\mu}{\sigma}
=\frac{1}{\sigma}(Q_\alpha X_\alpha - \mu)\,,
\qquad
\Delta D(\BX) 
= \Delta Q_{\alpha}\, X_{\alpha}\,.
\ee
Using the above expression for $z(\BX)$, 
it follows that the argument of the exponential can be written as
\be
\ba
-\frac{1}{2}z^2(\BX) 
&= -\frac{1}{2}\left(\frac{Q_\alpha Q_\beta}{\sigma^2}
X_\alpha X_\beta
-2\,\frac{\mu Q_\alpha}{\sigma^2}X_\alpha
+\frac{\mu^2}{\sigma^2}\right)
\\
&\equiv
-\frac{1}{2}\left(X_\alpha U_{\alpha\beta}X_\beta
-2V_\alpha X_\alpha+W\right)
\\
&=-\frac{1}{2}\left(\BX^T \BU \BX-2\BV^T\BX+W\right)\,,
\ea
\ee
where 
\be
U_{\alpha\beta} \equiv \frac{Q_\alpha Q_\beta}{\sigma^2}\,,
\qquad
V_\alpha \equiv \frac{\mu Q_\alpha}{\sigma^2}\,,
\qquad
W\equiv \frac{\mu^2}{\sigma^2}\,.
\label{e:UVW}
\ee
It is also useful to rewrite the combination of linear terms in 
$\Delta\mu$, $\Delta\sigma$, and $\Delta D(\BX)$ as
\be
\begin{aligned}
\Delta\mu+z(\BX)\,\Delta\sigma-\Delta D(\BX)
&=\Delta\mu
+\frac{\Delta\sigma}{\sigma}
\left(Q_\alpha X_\alpha-\mu\right)
-\Delta Q_\alpha X_\alpha
\equiv c+\BL^T\BX\,,
\end{aligned}
\ee
where
\be
c\equiv
\Delta\mu-\mu\,\Delta\sigma/\sigma\,,
\qquad
L_\alpha\equiv
Q_\alpha\,\Delta\sigma/\sigma
-\Delta Q_\alpha\,.
\label{e:c_L_defs}
\ee
These last two quantities encode all of the information in the $\Delta$ terms
connected with noise mismodeling.

Using the above results, 
we can first simplify the product of 
the Gaussian prefactor in \eqref{e:eps_def} and
the Gaussian measure in \eqref{e:ME_int2}
by completing the square of the combined argument.
After a few lines of algebra, we find
\be
\begin{aligned}
&{\rm exp}\left[-\frac{1}{2}z^2(\BX)\right]
{\rm exp}\left[-\frac{1}{2}(\BX-\BM)^T\BK^{-1}(\BX-\BM)\right]
\\
&\hspace{0.5in}
= {\rm exp}\left[-\frac{1}{2}\BX^T \BU\BX + \BV^T\BX - \frac{1}{2} W\right]
{\rm exp}\left[-\frac{1}{2}(\BX-\BM)^T\BK^{-1}(\BX-\BM)\right]
\\
&\hspace{0.5in}
= {\rm exp}\left[-\frac{1}{2}(\BX-\BA)^T\BB^{-1}(\BX-\BA)\right]
{\rm exp}\left[\frac{1}{2}\BA^T\BB^{-1}\BA - \frac{1}{2} \BM^T \BK^{-1}\BM - \frac{1}{2}W\right]\,,
\end{aligned}
\ee
where the matrix $\BB^{-1}$ and vector $\BA$ are defined by
\be
\begin{aligned}
B^{-1}_{\alpha\beta}
&\equiv K^{-1}_{\alpha\beta}+U_{\alpha\beta}\,,
\qquad
A_\alpha
\equiv B_{\alpha\beta}
\bigl(K^{-1}_{\beta\gamma}M_\gamma+V_\beta\bigr)\,,
\end{aligned}
\ee
with $B_{\alpha\beta}$ denoting the components of $\BB$, which is inverse
to the matrix $\BB^{-1}$.
The remaining Gaussian integral over $\BX$ has mean $\BA$ and covariance
$\BB$.  
Therefore,
\be
\begin{aligned}
&\int \dd \BX\,
\exp\left[-\frac{1}{2}(\BX-\BA)^T\BB^{-1}(\BX-\BA)\right]
\left(\Delta\mu + z(\BX)\Delta\sigma -\Delta D(\BX)\right)
\\
&\hspace{0.5in}
=\int \dd \BX\,
\exp\left[-\frac{1}{2}(\BX-\BA)^T\BB^{-1}(\BX-\BA)\right]
\left(c+\BL^T\BX\right)
\\
&\hspace{0.5in}=\sqrt{\det(2\pi \BB)}\left(c+\BL^T\BA\right)\,.
\end{aligned}
\ee
The final expression for the mean error is consequently
\be
{\rm ME}
=\frac{1}{\sqrt{2\pi}\sigma}
\frac{1}{\sqrt{\det(\BB^{-1}\BK)}}\,
\exp\left[
\frac{1}{2}\BA^T\BB^{-1}\BA
-\frac{1}{2}\BM^T\BK^{-1}\BM
-\frac{1}{2} W\right]
\left(c+\BL^T\BA\right)\,.
\label{e:ME_final2}
\ee
Since \eqref{e:eps_def} is the first-order change in the $p$-value,
\eqref{e:ME_final2} gives the mean error through order $\Delta$, with
corrections beginning at order $\Delta^2$.

To improve the estimate of the mean error for finite $\Delta$, we use
a trapezoidal endpoint approximation for $\epsilon(\BX)$ that averages
the values of the factors multiplying
$\Delta\mu$, $\Delta \sigma$, and $\Delta D(\BX)$ evaluated at the 
properly modeled and mismodeled values of $\BQ$ and $\BN$:
\be
\epsilon_{\trap}(\BX) 
\equiv A_{\trap}(\BX)\,\Delta\mu + B_{\trap}(\BX)\,\Delta\sigma 
+ C_{\trap}(\BX)\,\Delta D(\BX)\,,
\label{e:eps_mid_def}
\ee
where
\be
\ba
&A_{\trap}(\BX) \equiv\frac{1}{2}\left(
\frac{1}{\sqrt{2\pi}\sigma} e^{-\frac{1}{2}z^2(\BX)}
+\frac{1}{\sqrt{2\pi}\sigma_m} e^{-\frac{1}{2}z_m^2(\BX)}
\right)\,,
\\
&B_{\trap}(\BX) \equiv\frac{1}{2}\left(
\frac{1}{\sqrt{2\pi}\sigma} e^{-\frac{1}{2}z^2(\BX)}\,z(\BX)
+\frac{1}{\sqrt{2\pi}\sigma_m} e^{-\frac{1}{2}z_m^2(\BX)}\,z_m(\BX)
\right)\,,
\\
&
C_{\trap}(\BX) \equiv - A_{\trap}(\BX)\,.
\ea
\ee
By repeating the above calculation for the mean of
$\epsilon_{\trap}(\BX)$, we obtain
\be
{\rm ME}_{\trap}
\equiv \left\langle \epsilon_{\trap}(x)\right\rangle_{x\sim H_S}
= \frac{1}{2}({\rm ME} + {\rm ME}_m)\,,
\ee
%
where ${\rm ME}_m$ is given by \eqref{e:ME_final2} with 
$\BQ$ and $\BN$ replaced by their mismodeled values $\BQ_m$
and $\BN_m$.  The increments $\Delta\mu$, $\Delta\sigma$, and
$\Delta\BQ$ retain the mismodeled-minus-properly-modeled signs
defined in \eqref{e:Delta_convention}.

\subsection{Calculation of the mean-squared error}

The calculation of the mean-squared error:
\be
{\rm MSE}\equiv
\langle \epsilon^2(\BX)\rangle_{x\sim H_S}
\equiv 
\int \dd \BX \>
\frac{1}{\sqrt{{\rm det}(2 \pi \BK)}}
{\rm exp}\left[-\frac{1}{2}(X_\alpha-M_\alpha)
K^{-1}_{\alpha\beta}(X_\beta-M_\beta)\right]\,
\epsilon^2(\BX)
\label{e:MSE_int}
\ee
proceeds in a manner very similar to that of the mean error 
given in the previous subsection.
The two main differences involve the square of the $\Delta$ terms:
\be
\begin{aligned}
\left(\Delta\mu+z(\BX)\,\Delta\sigma-\Delta D(\BX)\right)^2
= c^2+ 2 c\,\BL^T\BX + (\BL^T\BX)^2\,,
\end{aligned}
\ee
and the square of the Gaussian pre-factor multiplying the 
Gaussian measure:
\be
\begin{aligned}
&{\rm exp}\left[-z^2(\BX)\right]
{\rm exp}\left[-\frac{1}{2}(\BX-\BM)^T\BK^{-1}(\BX-\BM)\right]
\\
&\hspace{0.5in}
= {\rm exp}\left[-\frac{1}{2}(\BX-\BG)^T \BH^{-1}(\BX-\BG)\right]
{\rm exp}\left[\frac{1}{2}\BG^T \BH^{-1} \BG - \frac{1}{2} \BM^T \BK^{-1}\BM - W\right]\,,
\end{aligned}
\ee
where
\be
\begin{aligned}
H^{-1}_{\alpha\beta}
&\equiv K^{-1}_{\alpha\beta}+2U_{\alpha\beta}\,,
\qquad
G_\alpha
\equiv H_{\alpha\beta}
\bigl(K^{-1}_{\beta\gamma}M_\gamma+2V_\beta\bigr)\,.
\end{aligned}
\ee
Together, these lead to
\be
\begin{aligned}
{\rm MSE}
=\frac{1}{2\pi\sigma^2}
\frac{1}{\sqrt{\det(\BH^{-1}\BK)}}\,
\exp\left[
\frac{1}{2}\BG^T\BH^{-1}\BG
-\frac{1}{2}\BM^T\BK^{-1}\BM-W\right]
\left(c^2+ 2c\,\BL^T\BG + \BL^T(\BH + \BG\BG^T)\BL\right)
\label{e:MSE_final2}
\end{aligned}
\ee
for the mean-squared error due to noise mismodeling.
Because it is obtained by squaring the first-order error,
\eqref{e:MSE_final2} is the leading, order-$\Delta^2$ contribution to
the MSE; corrections begin at order $\Delta^3$.

We can similarly calculate a trapezoidal endpoint-averaged version
of the mean-squared error by repeating the above 
calculation starting with $\epsilon_{\trap}(\BX)$ given in 
\eqref{e:eps_mid_def}.
By doing so, we obtain
\be
{\rm MSE}_{\trap}
\equiv \left\langle \epsilon^2_{\trap}(x)\right\rangle_{x\sim H_S}
= \frac{1}{4}({\rm MSE} + {\rm MSE}_m + 2{\rm MSE}_c)\,,
\label{e:MSE_mid}
\ee
where ${\rm MSE}_m$ is given by \eqref{e:MSE_final2} with 
$\BQ$ and $\BN$ replaced by their mismodeled values $\BQ_m$ and $\BN_m$,
while retaining the sign conventions of
\eqref{e:Delta_convention},
and the ``cross-term" ${\rm MSE}_c$ is given by
\be
\begin{aligned}
&{\rm MSE}_c
\equiv
\frac{1}{2\pi\sigma\,\sigma_m}
\frac{1}{\sqrt{\det\left(\bar\BH^{-1}\BK\right)}}\,
\exp\left[\frac{1}{2}\bar\BG^T \bar\BH^{-1} \bar\BG
-\frac{1}{2}\BM^T \BK^{-1} \BM-\bar W\right]
\\
&\hspace{1.5in}
\times
\left(c\,c_m+ (c\,\BL^T_m+c_m\BL^T) \bar\BG + \BL^T(\bar\BH + \bar\BG \bar\BG^T)\BL_m\right)\,.
\label{e:MSE_final3}
\end{aligned}
\ee
In the above expression:
\be
c_m\equiv
\Delta\mu-\mu_m\,\Delta\sigma/\sigma_m\,,
\qquad
L_{m,\alpha}\equiv
Q_{m,\alpha}\,\Delta\sigma/\sigma_m
-\Delta Q_\alpha\,,
\label{e:cm_Lm_defs}
\ee
and
\be
\bar \BH^{-1} = \BK^{-1} + 2\bar\BU\,,
\quad
\bar \BG = \bar\BH(\BK^{-1} \BM + 2\bar\BV)\,,
\ee
where
\be
\bar \BU\equiv \frac{1}{2}(\BU + \BU_m)\,,
\quad
\bar \BV\equiv \frac{1}{2}(\BV + \BV_m)\,,
\quad
\bar W\equiv \frac{1}{2}(W + W_m)\,,
\ee
with $\BU_m$, $\BV_m$, and $W_m$ being the mismodeled
versions of $\BU$, $\BV$, and $W$ in \eqref{e:UVW}.
}

\end{widetext}

\bibliographystyle{apsrev4-2-10authors}
\bibliography{references}

\end{document}